\documentclass
[aps,prb,endfloat,twocolumn,showpacs,preprintnumbers,amsmath,amssymb]{revtex4}
\usepackage{natbib}
\usepackage[dvips]{graphicx}
\usepackage{wrapfig}
\usepackage{dcolumn}
\usepackage{bm}
\usepackage{times}
\usepackage{mathptmx}
\usepackage{textcomp}
\usepackage[usenames]{color}
\usepackage{ulem}
\def\rfeo{$R$Fe$_2$O$_4$}

\begin{document}

\title{Magneto-Dielectric phenomena in charge and spin frustrated system of layered iron oxide}

\author{Makoto~Naka, Aya~Nagano$^{\ast}$, and Sumio~Ishihara}
  \affiliation{Department of Physics, Tohoku University, Sendai 980-8578, Japan}
\date{\today}

\begin{abstract}
Dielectric and magnetic phenomena in spin and charge frustrated system $R$Fe$_2$O$_4$ 
($R$ is a rare-earth metal ion) are studied. 
An electronic model for charge, spin and orbital degrees in a pair of triangular-lattice planes is derived. 
We analyze this model by utilizing the mean-field approximation and the Monte-Carlo simulation in a finite size cluster. 
A three fold-type charge ordered structure with charge imbalance between the planes is stabilized in finite temperatures. 
This polar charge order is reinforced by spin ordering of Fe ions. 
This novel magneto-dielectric phenomenon is caused by spin frustration and charge-spin coupling 
in the exchange interaction. 
We show cross-correlation effect in magnetic- and electric-field responses. 
Oxygen deficiency effect as an impurity effect in a frustrated charge-spin coupled system is also examined. 

\end{abstract}

\pacs{75.80.+q, 72.80.Ga, 75.10.-b, 77.80.-e}


\maketitle

\section{introduction}

Simultaneous existence of electric and magnetic polarizations and their control 
by external field have been recently revived as multiferroic 
phenomena in correlated electron oxides.~\cite{kimura03,hur04,saitoh95,cheong07} 
Behind large coupling between electric and magnetic moments, 
spin frustration plays dominant roles on multiferroic properties. 
Non-colinear spin structures, such as cycroid, spiral and so on, 
are realized on a frustrated geometry, 
and spontaneous electric polarization is induced 
to gain the symmetric/anti-symmetric exchange interactions.~\cite{katsura05,mostvoy06,sergienko06}
In this viewpoints, 
this class of materials are recognized as a spin driven ferroelectricity. 
Another class of ferroelectricity is possible 
in correlated electron systems. 
Apart from the integer filling of electron in valence bands, 
charge degree of freedom is active. 
In particular, around the quarter filling, 
the long-range charge order due to electron correlation 
is ubiquitously observed in several transition-metal compounds.~\cite{verwey39,tokura_rev,maekawa_rev} 
When electronic charge is ordered without inversion symmetry, 
a macroscopic electric polarization appears. 
This is a ferroelectricity driven by electronic charge degree of freedom. 
This class of ferroelectricity is realized in 
low-dimensional organic salts, such as the neutral-ionic transition system~\cite{tokura89,lemee97} 
and $\alpha$-(BEDT-TTF)$_2$I$_3$.~\cite{yamamoto} 
Charge polarized state observed in a layered structure manganite Pr(Sr$_{0.1}$Ca$_{0.9}$)$_2$Mn$_2$O$_7$
is attributed to the charge-orbital order associated with lattice distortion.~\cite{tokunaga06} 
A possibility of ferroelectricty in manganites is also proposed in theoretical viewpoint.~\cite{efremov04} 

Rare-earth iron oxides with layered crystal structure \rfeo \ ($R$=Lu, Y, Yb, Er)~\cite{kimizuka90} 
of the present interest 
belong to this class of ferroelectricity.
Crystal structure of \rfeo \ consists of paired Fe-O triangular-lattice layers and $R$-O blocks  
stacked along the $c$ axis. 
Schematic view of a paired Fe-O layer, termed the W-layer, is shown in Fig.~\ref{fig:wlayer}(a). 
Average valence of Fe ions is +2.5, 
implying that equal amounts of Fe$^{2+}$ and Fe$^{3+}$ occupy the W-layer. 
Therefore, this material is recognized as a spin-charge frustrated system. 
Charge structure was investigated by the electron and x-ray diffraction experiments.\cite{yamada00,yamada97,zhang07} 
In LuFe$_2$O$_4$, below 500K, streak-type diffuse scattering was observed 
along $(1/3\ 1/3\ l)$ lines, and below 320K, 
spots appear at $(1/3 \ 1/3 \ 3m+1/2)$ in the streak lines associated with 
zigzag modulations. 
In this paper, we use the hexagonal index, although the space group is $R{\bar 3}m$.  
These experimental results are interpreted as two- and three-dimensional charge orders of electrons. 
The three-dimensional order of Fe$^{2+}$ and Fe$^{3+}$ 
was also confirmed by the resonant x-ray scattering technique at Fe $K$-edge.~\cite{ikeda05} 
As for the magnetic properties, magnetization in LuFe$_2$O$_4$ starts to increase around 250K.~\cite{iida86} 
Neutron diffraction experiments revealed that 
magnetic Bragg peaks at $(1/3\ 1/3\ m)$ appear 
and a ferrimagnetic order realizes 
below this temperature.~\cite{akimitsu79,iida93,shiratori92,nagai07,christianson07} 
Electric polarization and dielectric anomalies were observed 
around the three-dimensional charge ordering temperature, 
although the dielectric constant shows strong dispersive and diffusive nature.~\cite{ikeda05,ikeda00} 
Several magneto-dielectric phenomena were also reported 
around the ferrimagnetic ordering temperature.~\cite{ikeda05,ikeda94a,subramanian06} 
It is worth noting that these dielectric and magnetic phenomena depend on the rare-earth metal element, $R$, 
and the oxygen stoichiometry; 
in YFe$_2$O$_4$, with decreasing temperature, 
the three fold-type charge order is changed into a four fold-type one which 
is extremely sensitive to oxygen deficiency.~\cite{funahashi84,ikeda02,ikeda03} 
These microscopic and macroscopic experiments denote that 
$3d$ electronic charges are responsible for the dielectric anomalies, 
and couple strongly with spins. 

To elucidate mechanism of dielectric phenomena in \rfeo ,  
Yamada and coworkers proposed a model for the three fold-type charge order.\cite{yamada97} 
This charge-structure model is shown in Fig.~\ref{fig:COs}(a) which 
will be introduced in more detail in Sect.~\ref{sect:charge}. 
This is a $\sqrt{3} \times \sqrt{3}$ structure in a plane,  
and along [110], 
electronic charges are aligned 
$\cdots$Fe$^{3+}$Fe$^{3+}$Fe$^{2+}$$\cdots$ in the lower plane and 
$\cdots$Fe$^{3+}$Fe$^{2+}$Fe$^{2+}$$\cdots$ in the upper one. 
That is, electronic charge is polarized between the upper and lower planes, 
and finite electric dipole moments exist in the W-layer. 
Based on this polar charge model and the neutron diffraction data, 
a possible spin structure in the ferrimagnetic ordered phase was proposed.\cite{ikeda02} 

A number of the experimental results~\cite{kimizuka90,yamada97,ikeda05,iida86,akimitsu79} 
and the analyses~\cite{yamada00,iida93,nagano1,nagano2,xiang07} 
imply that electronic processes and interactions 
are crucial for novel dielectric properties in this material. 
In this paper, we present a microscopic theory of electronic structure and 
magneto-dielectric phenomena in \rfeo. 
We focus on $3d$ electronic structure in a W-layer 
which is a minimum and main stage for the low-energy electronic state. 
We first suggest the orbital degree of freedom in a Fe$^{2+}$ ion, 
and derive an electronic Hamiltonian in a W-layer.  
This model consists of the long-range Coulomb interactions and the exchange 
interaction derived from the generalized $pd$ model. 
We analyze the charge structure by using the mean-field approximation and 
the Monte-Carlo (MC) simulation. 
The three-fold type polar charge order competes with other type non-polar ones, 
and is stabilized at finite temperature. 
This is caused by charge fluctuation in a frustrated triangular lattice. 
We furthermore examine spin structure and coupling between spin ordering and electric polarization. 
The polar charge order is strongly stabilized 
below the magnetic ordering temperature. 
This magneto-dielectric phenomenon is attributed to 
spin frustration in a triangular lattice. 
We demonstrate novel electric and magnetic responses 
which are available to examine the present theoretical scenario.  
Effects of oxygen deficiency on electric polarization are also studied. 

In Sect.~\ref{sect:model}, we derive the model Hamiltonian 
for electronic structure in a W-layer. 
In Sect.~\ref{sect:charge}, numerical results for 
the charge structure and electric polarization are presented. 
Calculated results for the spin structure and the magneto-dielectric responses 
are shown in Sect.~\ref{sect:spin}. 
Examined oxygen deficiency effects are introduced in Sect.~V. 
Section~\ref{sect:summary} is devoted to discussion and concluding remarks. 
Preliminary results for the present study 
have been published in Refs.~\onlinecite{nagano1} and~\onlinecite{nagano2}. 
Study of a doubly degenerate orbital model in a honeycomb lattice 
as an orbital model for \rfeo \ is presented in separate papers.~\cite{nagano2,nasu07} 

\section{model hamiltonian}
\label{sect:model}
\begin{figure}[t]
\includegraphics[width=\columnwidth]{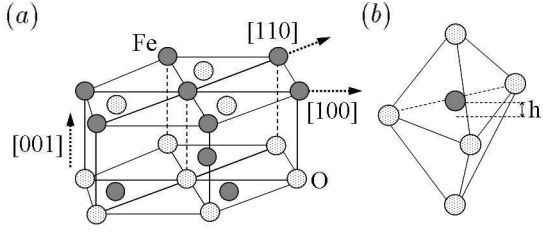}
\caption{
(a) : A pair of triangular-lattice planes (W-layer).
(b) : A FeO$_{5}$ cluster. 
}
\label{fig:wlayer}
\end{figure}

We start from the electronic structure 
in a single Fe ion in the W-layer.
This ion is five-fold coordinate 
with three O ions in the $xy$ plane and two at apices as shown 
in Fig.~\ref{fig:wlayer} (b).
We calculate the crystalline-field splitting 
of the Fe $3d$ orbitals in the FeO$_5$ cluster. 
Five O ions are replaced by point charges with valence of $-2e$, 
and their positions are determined by the crystal structure data.\cite{malaman75,kato76}
The hydrogen-like wave functions are adopted for the Fe $3d$ orbitals, 
and the effective nuclear charge is taken to be $+8$.
The split $3d$ orbitals are identified by the irreducible representation 
in the D$_{\rm 3d}$ group: 
the $d_{3z^{2}-r^{2}}$ orbital with $\rm A^{\prime}$, 
and two sets of the doubly degenerate orbitals 
$(-ad_{zx}+bd_{x^{2}-y^{2}}, \ ad_{yz}+bd_{xy})$ with $\rm E^{\prime}$,  
and $(ad_{x^{2}-y^{2}}+bd_{zx}, \ -ad_{xy}+bd_{yz})$ with $\rm E^{\prime\prime}$.
Numerical coefficients $a$ and $b$ satisfy the relation $a^{2}+b^{2}=1$.
We obtain that the degenerate $\rm E^{\prime}$ orbitals 
take the lowest energy with $b=0.89$, 
and the first excited level is $\rm E^{\prime\prime}$.
The energy difference between ${\rm E}^{\prime}$ and ${\rm E}^{\prime\prime}$, 
$\Delta E_{{\rm E}^{\prime}-{\rm E}^{\prime\prime}}$, is about 0.1eV 
which is smaller than that between $\rm E^{\prime\prime}$ and $\rm A^{\prime}$, 
$\Delta E_{\rm {\rm E}^{\prime\prime}-{\rm A}^{\prime}} \sim$ 0.6eV.
When we see the crystal structure in detail, 
an Fe ion is not located at center of a O$_5$ cage. 
Distance between the Fe ion and the O$_3$ plane denoted by $h$  
[see Fig.~\ref{fig:wlayer} (b)] 
is about 0.1$\rm \AA$ in LuFe$_2$O$_4$.\cite{malaman75,kato76}
We obtain that, with taking $h$ into account, 
$\Delta E_{{\rm E}^{\prime}-{\rm E}^{\prime \prime}}$ increases 
and $\Delta E_{{\rm E}^{\prime \prime}-{\rm A}^{\prime}}$ decreases.
The hybridization effects between Fe $3d$ and O $2p$ orbitals 
may increase these level separations. 
However, because of the small value 
of $\Delta E_{{\rm E}^{\prime}-{\rm E}^{\prime\prime}}$, 
we do not exclude a possibility that 
$(d_{x^{2}-y^{2}}, \ d_{xy})$ and $(d_{zx}, \ d_{yz})$  
couple strongly with each other, 
i.e. $a \sim b$, 
and that the ${\rm E}^{\prime \prime}$ level is the lowest.
In any cases, the lowest orbitals are degenerate.
As a result, in Fe$^{3+}$, 
each orbital is singly occupied, and total spin $S=5/2$ of the high-spin state. 
On the other hand, in Fe$^{2+}$, 
one of the degenerate lowest levels 
is doubly occupied by a hole, and $S=2$.
Thus, two fold orbital degeneracy exists in Fe$^{2+}$.
This is represented by the orbital pseudo-spin operator defined by 
\begin{equation}
{\bf T}_{i}=\frac{1}{2} \sum_{\mu \mu^{\prime} s} 
d_{i \mu s}^{\dagger} {\bf \sigma}_{\mu \mu^{\prime}} d_{i \mu^{\prime} s}, 
\label{eq:orbitalps}
\end{equation}
where $d_{i \mu s}^\dagger$ is the creation operator 
for an Fe $3d$ hole with orbital $\mu$, 
spin $s(=\uparrow, \ \downarrow)$ 
at site $i$, and ${\bf \sigma}_{\mu \mu^{\prime}}$ is the Pauli matrices.
In following part of this paper, 
we assume for simplicity that the two orbitals in the lowest level are $(d_{x^{2}-y^{2}}, d_{xy})$, 
and the index $\mu$ in Eq.~(\ref{eq:orbitalps}) takes the two. 
The $z$ component of the operator $T^{z}_{i}$ 
is $1/2$ ($-1/2$) for the state 
where a hole occupies the $d_{x^{2}-y^{2}}$ ($d_{xy}$) orbital. 
Even in the case where the orbitals in the lowest level 
are $(d_{zx}, d_{yz})$,  
the following part of this paper is valid 
by reinterpreting that the index $\mu$ in Eq.~(\ref{eq:orbitalps}) 
takes $(d_{yz}, \ d_{zx})$. 

We set up the model Hamiltonian for the electronic structure in a W-layer.
The $3d$ electrons in the Fe ions and the $2p$ ones in O 
which hybridizes with Fe $3d$ are introduced.
We start from the following generalized $pd$ Hamiltonian, 
\begin{equation}
{\cal H}_{pd} = {\cal H}_{d} + {\cal H}_{p} + {\cal H}_{t} + {\cal H}_{V} 
\label{eq : extended pd Hamiltonian},
\end{equation}
with 
\begin{align}
{\cal H}_{d} &=  \sum_{i \mu \sigma} \epsilon_{\mu}^{d} d_{i \mu \sigma}^\dagger d_{i \mu \sigma} 
\nonumber \\
&+ \sum_{i \mu} U^{d} n^{d}_{i \mu \uparrow} n^{d}_{i \mu \downarrow} 
+ \frac{1}{2} \sum_{i \mu \neq \mu^{\prime} \sigma \sigma^{\prime}} W^{d} n^{d}_{i \mu \sigma} n^{d}_{i \mu^{\prime} \sigma^{\prime}} 
\nonumber \\ 
& - \frac{1}{2} \sum_{i \mu \neq \mu^{\prime} \sigma  \sigma^{\prime}} I^{d} d_{i \mu \sigma}^{\dagger} d_{i \mu \sigma^{\prime}} d_{i \mu^{\prime} \sigma^{\prime}}^{\dagger} d_{i \mu^{\prime} \sigma},
\end{align}
\begin{align}
{\cal H}_{p} &=  \sum_{j \nu \sigma} \epsilon_{\nu}^{p} p_{j \nu \sigma}^\dagger p_{j \nu \sigma} 
\nonumber \\
&+ \sum_{j \nu} U^{p} n^{p}_{j \nu \uparrow} n^{p}_{j \nu \downarrow} 
+ \frac{1}{2} \sum_{j \nu \neq \nu^{\prime} \sigma  \sigma^{\prime}} W^{p} n^{p}_{j \nu \sigma} n^{p}_{j \nu^{\prime} \sigma^{\prime}} \nonumber \\
& - \frac{1}{2} \sum_{j \nu \neq \nu^{\prime} \sigma \sigma^{\prime}} I^{p} p_{j \nu \sigma}^{\dagger} p_{j \nu \sigma^{\prime}} p_{j \nu^{\prime} \sigma^{\prime}}^{\dagger} p_{j \nu^{\prime} \sigma},
\end{align}
\begin{equation}
{\cal H}_{t} = \sum_{i \eta \sigma} 
t^{pd} d_{i \eta_{x}^{2}-\eta_{y}^{2} \sigma}^{\dagger} p_{i+\delta_\eta \eta_{y} \sigma} + {\rm H.c.},
\end{equation}
\begin{align}
{\cal H}_{V} = \left ( \sum_{\langle ij \rangle}^{\rm abNN} V_{\rm abNN}  
+ \sum_{\langle ij \rangle}^{\rm cNN} V_{\rm cNN} 
+ \sum_{\langle ij \rangle}^{\rm cNNN} V_{\rm cNNN}  \right ) n^{d}_{i} n^{d}_{j}, 
\end{align}
where $d_{i \mu \sigma}^\dagger$ is the creation operator 
for the Fe $3d$ hole with orbital $\mu$ 
($=xy,~x^{2}-y^{2},~yz,~zx,~3z^{2}-r^{2}$) 
and spin $\sigma$ ($=\uparrow, \downarrow)$ 
at site $i$, and 
$p_{j \nu \sigma}^\dagger$ is 
for the O $2p$ hole with orbital $\nu$ 
($=x,~y,~z$). 
Number operators are defined by 
$n^{d}_{i \mu \sigma}=d_{i \mu \sigma}^\dagger d_{i \mu \sigma}$, 
$n^{p}_{j \nu \sigma}=p_{j \nu \sigma}^\dagger p_{j \nu \sigma}$, 
and 
$n_i^d=\sum_{\mu \sigma} n^d_{i \mu \sigma}$. 
A simbol $\delta_\eta$ is a connecting vector between Fe and NN O ions along directin $\eta$. 
%
\begin{figure}
\includegraphics[width=0.5\columnwidth]{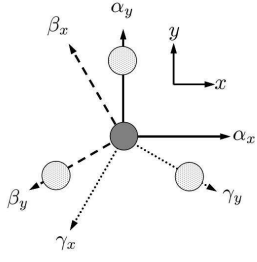}
\caption{
Three two-dimensional coordinates, $(\alpha_{x}, \alpha_{y})$, $(\beta_{x}, \beta_{y})$ 
and $(\gamma_{x}, \gamma_{y})$ in a FeO triangular lattice.
Filled and dotted circles represent Fe and O ions, respectively.
}
\label{fig:rotate}
\end{figure}
Interactions in a Fe ion are described in 
the 1st term of Eq.~(\ref{eq : extended pd Hamiltonian}), ${\cal H}_{d}$, 
where the level energy $\epsilon_{\mu}^{d}$, 
the intra-orbital Coulomb interaction $U^{d}$, 
the inter-orbital one $W^{d}$, 
and the exchange interaction $I^{d }$ are considered.
Interactions in ${\cal H}_{p}$ are defined in the same way with 
those in ${\cal H}_{d}$.
Hopping of a hole between the nearest neighboring (NN) Fe and O ions 
in the same plane is described in ${\cal H}_{t}$ with the transfer integral $t^{pd}$.
For convenience, we introduce the three two-dimensional coordinates $(\eta_{x}, \eta_{y})$ 
with $\eta=(\alpha, \beta, \gamma)$, which are obtained 
by a rotation of the $xy$ axis by $2 \pi m_{\eta}/3$ 
with $(m_{\alpha}, m_{\beta}, m_{\gamma})=(0, 1, 2)$ (see Fig.~\ref{fig:rotate}).
In each coordinate, we define the operators as 
\begin{equation}
\begin{pmatrix}
d_{i \eta_{x}^{2}-\eta_{y}^{2} \sigma} \\ 
d_{i \eta_{x} \eta_{y} \sigma} 
\end{pmatrix}
=
\begin{pmatrix}
\cos{\frac{4 \pi}{3} m_{\eta}} & \sin{\frac{4 \pi}{3} m_{\eta}} \\ 
-\sin{\frac{4 \pi}{3} m_{\eta}} & \cos{\frac{4 \pi}{3} m_{\eta}} 
\end{pmatrix}
\begin{pmatrix}
d_{i x^{2}-y^{2} \sigma} \\ 
d_{i xy \sigma}
\end{pmatrix},
\end{equation}
and 
\begin{equation}
\begin{pmatrix}
p_{i \eta_{x} \sigma} \\ 
p_{i \eta_{y} \sigma} 
\end{pmatrix}
=
\begin{pmatrix}
 \cos{\frac{ 2 \pi}{3} m_{\eta}} & \sin{\frac{2 \pi}{3} m_{\eta}} \\ 
-\sin{\frac{ 2 \pi}{3} m_{\eta}} & \cos{\frac{2  \pi}{3} m_{\eta}}
\end{pmatrix}
\begin{pmatrix}
p_{i x \sigma} \\ 
p_{i y \sigma}
\end{pmatrix}.
\end{equation}
In the bond direction $\eta$, 
the $d_{\eta_{x}^{2}-\eta_{y}^{2}}$ and $p_{\eta_{y}}$ orbitals compose 
the $\sigma$ bond.
%
\begin{figure}
\centering
\includegraphics[width=0.5\columnwidth]{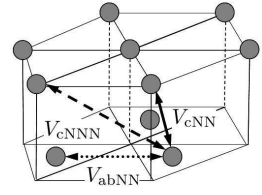}
\caption{
Inter-site Coulomb interactions between Fe ions.
Solid, broken and dotted arrows represent interactions 
between the nearest neighbor ($V_{\rm cNN}$), 
the next nearest neighbor ($V_{\rm abNN}$) 
and the third neighbor ($V_{\rm cNNN}$) Fe-Fe bonds, 
respectively.
}
\label{fig:LRC}
\end{figure}
The inter-site Coulomb interactions between Fe ions 
are taken into account in the last term of Eq.~(\ref{eq : extended pd Hamiltonian}), i.e. ${\cal H}_{V}$.
We consider the largest three interactions in the W-layer 
as shown in Fig.~\ref{fig:LRC}: 
the inter-plane NN interaction ($V_{\rm cNN}$), 
the intra-plane NN one ($V_{\rm abNN}$) 
and the inter-plane next NN one ($V_{\rm cNNN}$). 
This is because (1) these Coulomb interactions are 
a minimum set which reproduces the three-types of charge structures 
observed experimentally, 
and (2) a distance between the 4th neighbor Fe ions 
in the W-layer is comparable to that between the NN W-layers. 
This will be discussed in Sect.~\ref{sect:charge} in more detail. 
Summations in ${\cal H}_V$ take the three kinds of pairs. 
When the $1/r$-type Coulomb interaction is assumed, 
we obtain $V_{\rm cNN}/V_{\rm abNN}=1.2$ and $V_{\rm cNNN}/V_{\rm abNN}=0.77$ 
for LuFe$_{2}$O$_{4}$. 
By introducing the pseudo-spin operator $Q_{i}^{z}$ 
for charge degree of freedom, 
${\cal H}_{V}$ is rewritten as an antiferromagnetic Ising model 
\begin{equation}
{\cal H}_{V}=\left ( \sum_{\langle ij \rangle}^{\rm cNN} V_{\rm cNN}  
+ \sum_{\langle ij \rangle}^{\rm abNN} V_{\rm abNN}  
+ \sum_{\langle ij \rangle}^{\rm cNNN} V_{\rm cNNN} \right ) Q_{i}^{z} Q_{j}^{z} , 
\label{eq:hv}
\end{equation} 
where a constant term is omitted.  The operator $Q_i^z$ takes 
$1/2$ and $-1/2$ for Fe$^{3+}$ and Fe$^{2+}$, respectively. 
The charge conservation is imposed by a relation $\sum_{i} Q_{i}^{z}=0$.

Based on the extended $pd$ Hamiltonian ${\cal H}_{pd}$, we derive the effective Hamiltonian  
for the superexchange interactions between NN Fe ions in a plane.
This interaction arises from virtual hopping of holes 
between Fe ions.
The Hamiltonian is derived by the 4th order projection-perturbation procedure 
in terms of the hopping term ${\cal H}_{pd}$.
Following two exchange processes 
are considered: 
\begin{eqnarray}
d^{M}p^{0}d^{N} \to d^{M+1}p^{0}d^{N-1} \to d^{M}p^{0}d^{N} , 
\label{eq:dd}
\end{eqnarray}
and 
\begin{eqnarray}
d^{M}p^{0}d^{N} \to d^{M-1}p^{2}d^{N-1} \to d^{M}p^{0}d^{N} , 
\label{eq:dpd}
\end{eqnarray}
where we adopt the hole picture, and $M$ and $N$ represent the numbers of holes. 
These are termed the $dd$- and $dpd$-processes, respectively, from now on. 
Here, we present the outline of derivation, 
and details are given in Appendix A.
A general form of the Hamiltonian is 
\begin{equation}
{\cal H}_{J} = 
\hat{P} {\cal H}_{t} \frac{1}{E_{i} - {\cal H}_{0}} 
\hat{Q} {\cal H}_{t} \frac{1}{E_{i} - {\cal H}_{0}} \hat{Q} {\cal H}_{t} \frac{1}{E_{i} - {\cal H}_{0}} 
\hat{Q} {\cal H}_{t} \hat{P}, 
\label{eq:perturbation}
\end{equation}
where $\hat{P}$ is the projection operator 
for the $d^{5}$ ($d^{4}$) high-spin states in Fe$^{3+}$ (Fe$^{2+}$), 
$\hat{Q}=1-\hat{P}$, 
and $E_i$ is the initial-state energy. 
Many body effects of ${\cal H}_V$ in the intermediate states 
are considered approximately;  
we assume that the intermediate-state energies for $d^{M+1}p^{0}d^{N-1}$ 
[see Eq.(\ref{eq:dd})] are higher than the initial- and final-state energies  
by a constant energy parameter $\widehat V$, which is of the order of the 
inter-site Coulomb interaction. 
Then, we set ${\cal H}_{0}={\cal H}_{d} + {\cal H}_{p} + \widehat V$. 
We interpret that effects of ${\cal H}_V$ in other intermediate states 
are included in the charge-transfer energy, $\Delta_{\rm CT}$, between 
the $d_{\eta^2_{x}-\eta^{2}_{y}}$ and $p_{\eta_{y}}$ orbitals. 
The obtained Hamiltonian ${\cal H}_J$ 
is classified by valences of Fe ions in the initial states, 
i.e. Fe$^{m+}$-Fe$^{n+}$ where $n$ and $m$ take $2$ and $3$, 
and electron configurations in the intermediate states denoted by $k$. 
The Hamiltonian is given as 
\begin{equation}
{\cal H}_{J} = \sum_{(mn) \ k} {\cal H}^{(mn)-k}.
\label{eq:hj}
\end{equation}
All possible intermediate states $k$ are taken into account in Eq.~(\ref{eq:hj}) 
which consists of 6 terms in ${\cal H}^{(22)-k}$ and ${\cal H}^{(23)-k}$, 
and 4 terms in ${\cal H}^{(33)-k}$. 
Explicit formulae of all terms are presented in Appendix A.
Here, we show some representative terms: 
\begin{align}
{\cal H}^{(22)-1} &= J^{(22)-1} \sum_{\langle ij \rangle} 
\left ( {\bf I}_{i} \cdot {\bf I}_{j} + 6 \right) 
\left(\frac{1}{2} - 2 \tau_{i \eta_i} \tau_{j \eta_j} \right) 
\nonumber \\
& \times
\left(\frac{1}{2} - Q_{i}^{z} \right) \left(\frac{1}{2} - Q_{j}^{z} \right), 
\end{align}
\begin{align}
{\cal H}^{(23)-1} &= J^{(23)-1} \sum_{\langle ij \rangle} 
\left({\bf J}_{i} \cdot {\bf I}_{j} + \frac{15}{2} \right) 
\left(\frac{1}{2} - \tau_{j \eta_j} \right)
\nonumber \\
& \times
\left(\frac{1}{2} + Q_{i}^{z} \right) \left(\frac{1}{2} - Q_{j}^{z}\right), 
\end{align}
\begin{align}
{\cal H}^{(33)-1} &= J^{(33)-1} \sum_{\langle ij \rangle} 
\left({\bf J}_{i} \cdot {\bf J}_{j} - \frac{25}{4} \right)
\nonumber \\
& \times
\left(\frac{1}{2} + Q_{i}^{z} \right) 
\left(\frac{1}{2} + Q_{j}^{z} \right).
\end{align}
We define the spin operators ${\bf I}_{i}$ and ${\bf J}_{i}$ 
for Fe$^{2+}$ and Fe$^{3+}$ with amplitudes of $2$ and $5/2$, 
respectively.
The orbital operator is redefined 
in the $(\eta_{x} , \eta_{y})$ coordinate as 
\begin{equation}
\tau_{i \eta} = T_{i}^{z} \cos \left ( {\frac{2 \pi}{3} m_{\eta}} \right )
+ T_{i}^{x} \sin \left ( {\frac{2 \pi}{3} m_{\eta}}  \right )  . 
\end{equation}
This operator takes $1/2$ ($-1/2$), 
when the $d_{\eta^{2}_{x}-\eta^{2}_{y}}$ ($d_{\eta_{x} \eta_{y}}$) orbital  
is occupied by a hole. 
In a given pair of $i$ and $j$ sites, 
subscripts $\eta_i$ and $\eta_j$ in $\tau_{i \eta_i}$ and $\tau_{j \eta_j}$ are 
automatically determined. 
The exchange constants are defined by  
$J^{(22)-1}=- t_{dd{\rm c}}^2/[10 \Delta_{(22)-1}] $, 
$J^{(23)-1}=-2t_{dd{\rm c}}^2/[25 \Delta_{(23)-1}] $,  and 
$J^{(33)-1}=4t_{dd{\rm c}}^2/[25 \Delta_{(33)-1}]$ 
where $t_{dd{\rm c}}$ is the transfer integral between NN Fe ions 
defined by 
$t_{dd{\rm c}}=(t_{pd}^2 \cos \theta)/\Delta_{\rm CT}$ 
with the Fe-O-Fe bond angle $\theta(=120^\circ)$. 
We introduce the intermediate-state energies as 
$\Delta_{(22)-1}=W^{d}-I^{d}+{\widehat V}$, 
$\Delta_{(23)-1}={\widehat V}$, and 
$\Delta_{(33)-1}=U^{d} + 4I^{d}+{\widehat V}$. 
It is worth to note that 
1)  ${\cal H}^{(22)-l}$ is expressed as a product of charge, 
spin and orbital interactions between given sites $i$ and $j$, and 
2) ${\cal H}^{(32)-l}$ includes a linear term of the orbital pseudo spin 
because Fe$^{3+}$ dose not have the orbital degree of freedom.

\begin{figure}[t]
\centering
\includegraphics[width=\columnwidth]{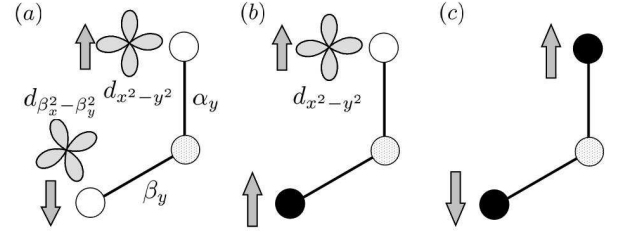}
\caption{
The lowest-energy spin and orbital configurations for a Fe$^{2+}$-Fe$^{2+}$ bond (a), 
for Fe$^{3+}$-Fe$^{2+}$ (b), and for Fe$^{3+}$-Fe$^{3+}$ (c). 
Open, filled and dotted circles represent Fe$^{2+}$, Fe$^{3+}$ and O ions, respectively. 
Spin and orbital configuration in (a) is energetically close to 
ferromagnetic spin alignment with $d_{x^2-y^2}$ and $d_{\beta_x \beta_y}$ orbitals. 
}
\label{fig:exchange}
\end{figure}
After all, we obtain the Coulomb- and exchange-interaction Hamiltonian 
\begin{equation}
{\cal H}={\cal H}_{V}+{\cal H}_{J}, 
\label{eq : effective Hamiltonian}
\end{equation}
where ${\cal H}_V$ and ${\cal H}_J$ are given in Eqs.~(\ref{eq:hv}) and (\ref{eq:hj}), respectively.  
Before going to the numerical results calculated in the Hamiltonian, 
we briefly mention the energy scales of charge, spin and orbital degrees of freedom, and 
signs of the exchange interactions.  
The inter-site Coulomb interactions provide a larger energy scale than the exchange interactions. 
Thus, the charge sector is frozen at the highest temperature in comparison with spin and orbital ones. 
This is consistent with the experimental results in LuFe$_{2}$O$_{4}$ 
where the charge ordering temperature (about 320K) is higher than 
the spin ordering one (about 250K).\cite{yamada00,yamada97,iida93}
By calculating the exchange energy in a given NN bond, 
we estimate stable spin and orbital configurations. 
This is not trivial from the Goodenough-Kanamori rule because of the 120$^\circ$ bond angle.  
The energy parameter sets for the exchange coupling constant $J^{(mn)-l}$ 
are determined from the experimental data in LaFeO$_3$,\cite{mizokawa96,mizuno98} 
and ${\bf I}_i$ and ${\bf J}_i$ are assumed to be Ising spins.  
We obtain the spin and orbital configurations for the lowest exchange energies as  
1) for a Fe$^{2+}$-Fe$^{2+}$ bond, $I_i^z I_j^z=-4$ (antiferromagnetic) and $\tau_i=\tau_j=1/2$,
which is energetically close to $I_i^z I_j^z=4$ (ferromagnetic) and $\tau_i=-\tau_j=1/2$, 
2) for Fe$^{3+}$-Fe$^{3+}$, $J_i^z J_j^z=-25/4$ (antiferromagnetic), and 
3) for Fe$^{2+}$-Fe$^{3+}$, $I_i^z J_j^z=5$ (ferromagnetic) and $\tau_i=1/2$. 
Schematic views for the stable configurations are presented in Fig.~\ref{fig:exchange}.  
In the neutron scattering experiments, 
the ferrimagnetic phase indexed as $(1/3 \ 1/3 \ m)$ appears around 250K. 
Possible magnetic structures are shown in Fig.~\ref{fig:spin3}, 
which will be explained in more detail later. 
%
In this structure, 
Fe$^{2+}$ ions in the 2Fe$^{3+}$-Fe$^{2+}$ (upper) plane are surrounded by six NN Fe$^{3+}$. 
Thus, the exchange Hamiltonian in this plane is 
reduced into a form of 
$
\sum_{\langle ij \rangle} [(1/2) \pm \tau_{i \eta_i} ]
$
which becomes a constant by using the relation of $\sum_{\langle ij \rangle} \tau_{i \eta_i}=0$.
This relation is also applicable to the Fe$^{2+}$-Fe$^{3+}$ bonds in the 
2Fe$^{2+}$-Fe$^{3+}$ (lower) plane  
where three Fe$^{2+}$-Fe$^{3+}$ bonds connecting a Fe$^{2+}$ ion are equivalent [see Fig.~\ref{fig:spin3}]. 
Therefore, the orbital part of the exchange Hamiltonian in this ferrimagnetic phase 
is mapped onto the following orbital model defined on a Fe$^{2+}$ sublattice; 
\begin{equation}
{\cal H}_{\rm orb}=J_{\rm orb}\sum_{i}^{\prime} 
\left ( \tau_{i \beta} \tau_{i+{\bf e}_\alpha \gamma}
+\tau_{i \gamma} \tau_{i+{\bf e}_\beta \alpha}  
+\tau_{i \alpha} \tau_{i+{\bf e}_\gamma \beta} \right ) , 
\end{equation}
where (${\bf e}_\alpha, {\bf e}_\beta, {\bf e}_\gamma$) represent the 
three unit vectors connecting NN Fe$^{2+}$ sites in 
a honeycomb lattice. 
A summation $\sum_{i}^{\prime}$ takes Fe$^{2+}$ sites in one of the two sublattices in 
a honeycomb lattice.
The coupling constant $J_{\rm orb}$ is given by 
the exchange constants $J^{(nm)-k}$.
In this model, it is obtained theoretically 
that the orbital does not show a conventional long-range order down to 
very low temperature of the order of 0.005$J_{\rm orb}$. 
Therefore, for simplicity, 
we assume that the pseudo-spin operators for orbital in ${\cal H}_J$ are set to be zero in the following calculation.
Theoretical study of the orbital model on a honeycomb lattice 
is presented in separate papers.~\cite{nagano2,nasu07} 

\section{Charge structure and electric polarization}
\label{sect:charge}
%
\begin{figure}[t]
\centering
\includegraphics[width=0.8\columnwidth]{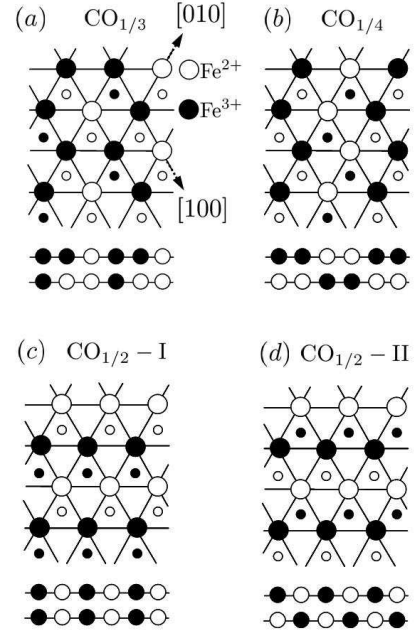}
\caption{
Charge structures in a W-layer : 
(a) CO$_{1/3}$, (b) CO$_{1/4}$, (c) CO$_{1/2}$-I,  
and (d) CO$_{1/2}$-II.
Filled and open circles represent Fe$^{3+}$ and Fe$^{2+}$, 
respectively, and 
large and small circles are for Fe ions in the upper 
and lower planes, respectively.
Lower panel in each figure is a side view from $[1{\bar 1}0]$.
}
\label{fig:COs}
\end{figure}
First we focus on the charge structure and the electric polarization 
by analyzing the inter-site Coulomb interaction term ${\cal H}_{V}$.
We apply, at the first stage, the mean-field approximation 
to ${\cal H}_{V}$, and obtain stable charge structures. 
The charge conservation is taken into account 
by adding the chemical potential term, 
$-V_{\rm ext}\sum_i Q_i^z $, in the Hamiltonian. 
We assume that the expectation value $\langle Q_{i}^z \rangle$ is periodic 
along the $\langle 110 \rangle$ or $\langle 210 \rangle$ directions, 
and takes the same amplitude along $\langle {\bar 1}10 \rangle$ 
or $\langle 010 \rangle$, respectively. 
Periodicity $L$ is taken up to $12$.
In upper and lower planes, $\langle Q_{i}^z \rangle$'s 
are independent and have the same periodicity along 
the $\langle 110 \rangle$ or $\langle 210 \rangle$ directions.
Each solution is characterized by the momentum 
${\bf q} \equiv (M/2L, M/2L, n)$ or $(M/2L, 0, n)$ 
where $M$ is the number of nodes of $\langle Q_{i}^z \rangle$ 
along the $\langle 110 \rangle$ or $\langle 210 \rangle$ directions, 
respectively. 
When a phase difference 
between $\langle Q_{i}^z \rangle$'s in the upper and lower planes is $0$ ($\pi$), 
$n$ takes $0$ ($1/2$).
Phase diagram is determined by comparing the free energy.
Representative charge structures are shown 
in Fig.~\ref{fig:COs}.
Four types of CO's in this figure, 
denoted by CO$_{1/3}$, CO$_{1/4}$, 
CO$_{1/2}\textrm{-I}$ and CO$_{1/2}\textrm{-II}$, 
are characterized by momenta 
${\bf q}=(1/3,1/3,0) \equiv {\bf q}_{1/3}$, 
$(1/4,1/4,1/2) \equiv {\bf q}_{1/4}$, 
$(1/2,1/2,0) \equiv {\bf q}_{1/2\textrm{-I}}$ 
and $(1/2,0,0) \equiv {\bf q}_{1/2\textrm{-II}}$, respectively.
As suggested by Yamada and coworkers,
CO$_{1/3}$ shows finite electric polarization due to charge imbalance 
between the two triangular-lattice planes.~\cite{yamada00,yamada97}
A ratio of Fe$^{2+}$ and Fe$^{3+}$ is 
$1:2$ ($2:1$) in the upper (lower) plane. 
In other charge structures, 
equal numbers of Fe$^{2+}$ and Fe$^{3+}$
occupy the upper and lower planes, 
and there is no electric polarization. 
%
\begin{figure}[t]
\centering
\includegraphics[width=0.8\columnwidth]{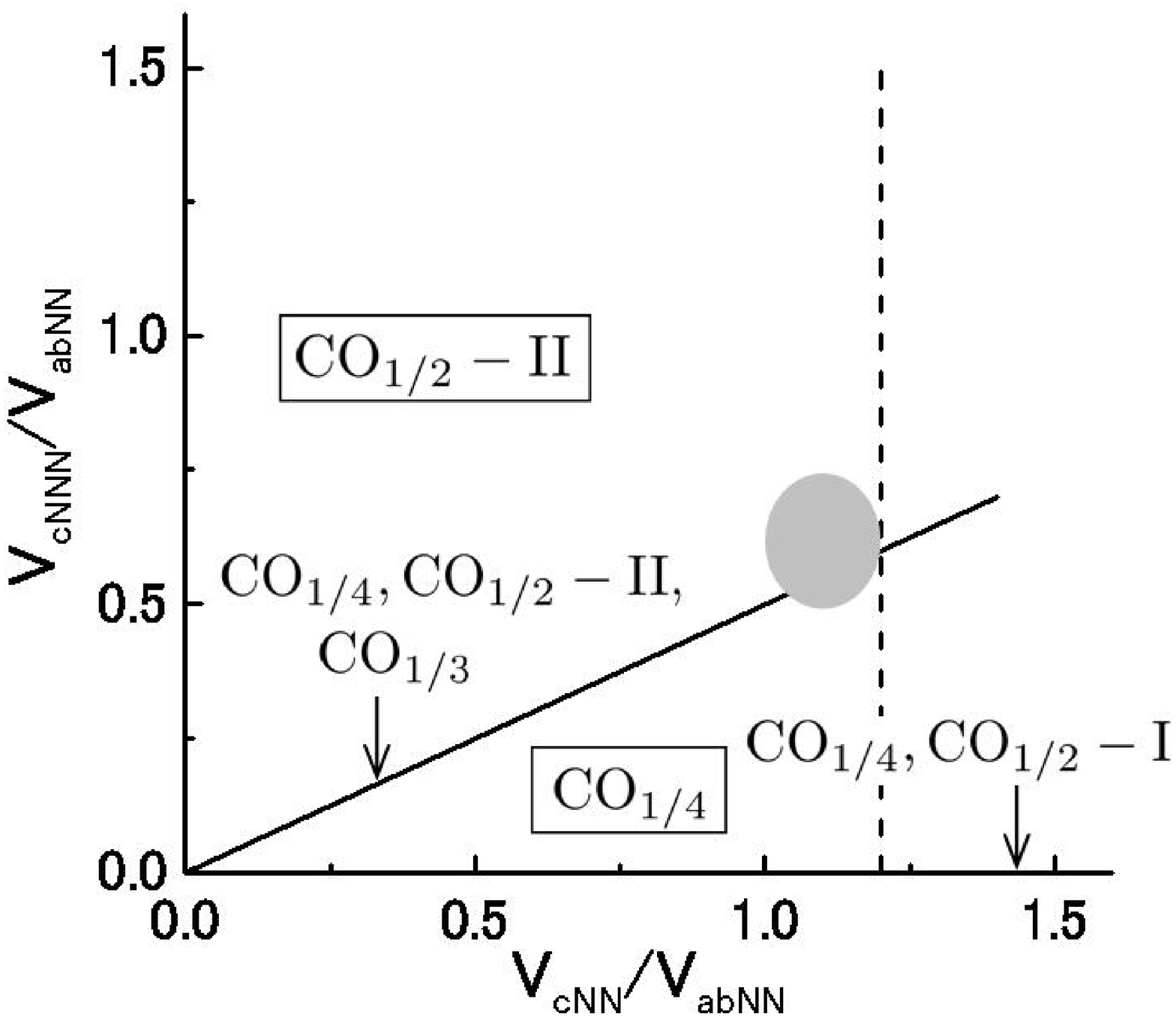}
\caption{
Mean-field phase diagram for charge order at zero temperature.
Three charge structures,  
CO$_{1/4}$, CO$_{1/2}\textrm{-II}$, and CO$_{1/3}$, are degenerate 
on a line of $V_{\rm cNN}=2V_{\rm cNNN}$, 
and the two, CO$_{1/4}$ and CO$_{1/2}\textrm{-I}$, are degenerate 
on a line of $V_{\rm cNNN}=0$.
Shaded area corresponds to a region for $R$Fe$_{2}$O$_{4}$.
Phase diagram in finite temperatures, shown in Fig.~\ref{fig:COphaseT}, 
is calculated on broken line.
}
\label{fig:COphase0}
\end{figure}

\begin{figure}[t]
\centering
\includegraphics[width=0.85\columnwidth]{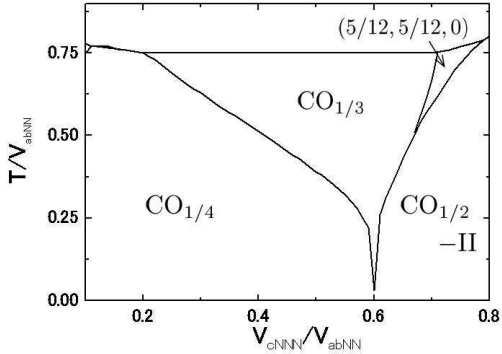}
\caption{
Mean-field phase diagram for charge structure in finite temperatures. 
The Coulomb interaction $V_{\rm cNN}/V_{\rm abNN}$ is chosen to be 1.2.
}
\label{fig:COphaseT}
\end{figure}

Mean-field phase diagram at zero temperature is presented in Fig.~\ref{fig:COphase0}. 
The non-polar CO$_{1/2}\textrm{-II}$ and CO$_{1/4}$ structures are stable 
in the regions of $V_{\rm CNN}/V_{\rm CNNN}<2$ and $V_{\rm CNN}/V_{\rm CNNN}>2$, respectively. 
The polar CO$_{1/3}$ structure appears only on the phase boundary 
where CO$_{1/3}$ is degenerate with CO$_{1/2}\textrm{-II}$ and CO$_{1/4}$.
Realistic parameter values for $R$Fe$_{2}$O$_{4}$ 
correspond to a shaded area in Fig.~\ref{fig:COphase0}. 
We fix a value of $V_{\rm cNN}/V_{\rm abNN}$ to be 1.2, as shown by a dashed line 
in Fig.~\ref{fig:COphase0},  
and calculate finite-temperature phase diagram (Fig.~\ref{fig:COphaseT}). 
The polar CO$_{1/3}$ is 
stabilized in a wide region 
between CO$_{1/4}$ and the CO$_{1/2}\textrm{-II}$.

Beyond the mean-field calculation, we examine 
the charge structure in finite temperature 
by using the MC simulation.
To avoid a trap of a simulation in local minima, 
we adopt the multi-canonical MC (MUMC) method.~\cite{berg96} 
Simulations are performed on a paired triangular lattice of 
$L \times L \times 2 (\equiv 2N)$ ($L=6$ and $12$) sites with the periodic-boundary condition 
in the $ab$ plane.
We use $6 \times 10^{6}$ MC steps to obtain a histogram in the MUMC method 
and $16 \times 10^{6}$ MC steps for measurement.
We calculate the charge correlation function 
and the electric polarization $P$ defined by 
\begin{equation}
N({\bf q}) = \frac{1}{(2N)^2} \sum_{ij} \langle Q_{i}^z Q_{j}^z \rangle e^{-i {\bf q} \cdot ({\bf r}_{i}-{\bf r}_{j})},
\end{equation}
\begin{equation}
P = \langle  p^2 \rangle^{1/2}. 
\end{equation}
with  
\begin{equation}
p = \frac{1}{N}\left ( \sum_{i}^{u}  - \sum_{i}^{l} \right )  Q_{i}^z , 
\end{equation}
where 
$r_i$ is a position of site $i$,   
and $\sum_{i}^{u(l)}$ represents a summation of site $i$ 
in the upper (lower) plane.
%
\begin{figure}
\centering
\includegraphics[width=0.75\columnwidth]{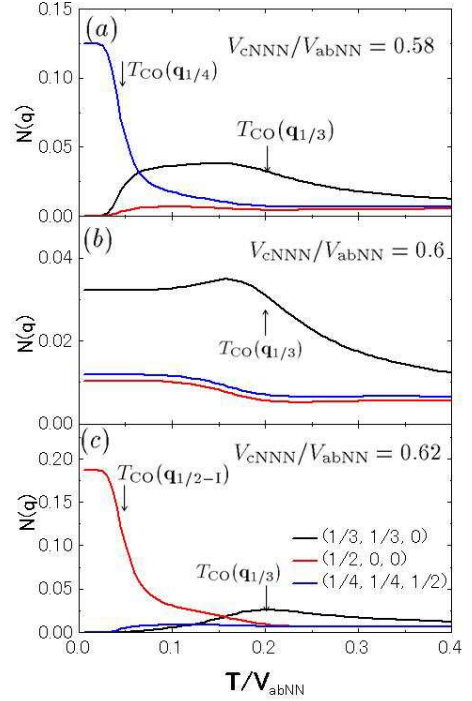}
\caption{Charge correlation functions $N({\bf q})$ 
at $V_{\rm cNNN}/V_{\rm abNN}=0.60$ (a), 
$0.58$ (b) 
and $0.62$ (c) calculated in ${\cal H}_V$.
The Coulomb interaction $V_{\rm cNN}/V_{\rm abNN}$ 
is chosen to be 1.2. 
}
\label{fig:cchv}
\end{figure}
\begin{figure}
\centering
\includegraphics[width=0.8\columnwidth]{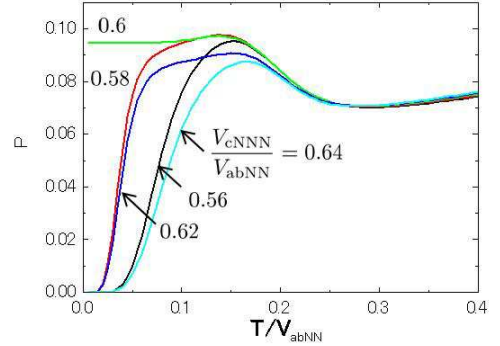}
\caption{
Electric polarization $P$ 
as a functions of $V_{\rm cNNN}/V_{\rm abNN}$ calculated in ${\cal H}_V$.
The Coulomb interaction $V_{\rm cNN}/V_{\rm abNN}$ 
is chosen to be 1.2.
}
\label{fig:polarization}
\end{figure}
\begin{figure}[t]
\includegraphics[width=0.4\columnwidth]{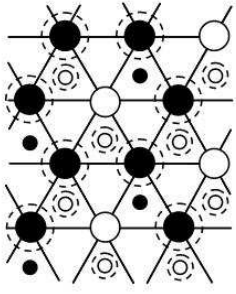}
\caption{
Two-sublattice structure in the CO$_{1/3}$ phase. 
Amplitude of the mean-field at Fe sites surrounded by broken circles 
is 0.9$V_{\rm abNN}$ and that at other sites is 2.1$V_{\rm abNN}$.}
\label{fig:MF1/3}
\end{figure}

The charge correlation functions  
at $V_{\rm cNNN}/V_{\rm abNN}=0.58$, $0.6$ and $0.62$ 
are presented in Fig.~\ref{fig:cchv}.
At $V_{\rm cNNN}/V_{\rm abNN}=0.6$,  
$N({\bf q}_{1/3})$ shows a hump around $\widetilde T \equiv T/V_{\rm abNN}=0.18$ 
and keeps a finite value down to the low temperature limit.
From the specific heat data, 
we identify $\widetilde T=0.18$ corresponds to the charge-ordering temperature. 
In the both cases of $V_{\rm cNNN}/V_{\rm abNN}=0.58$ and $0.62$, 
$N({\bf q}_{1/3})$ is dominant in high temperatures,  
starts to decrease around $\widetilde T=0.2$ and disappears at the lowest temperature.
On the contrary, the charge correlation $N({\bf q}_{1/2-{\rm II}})$ and $N({\bf q}_{1/4})$ 
grow up around $\widetilde T=0.05$, and  
increase with decreasing temperature.
These results and the specific heat data imply that 
the charge order at ${\bf q}_{1/3}$ is changed into the other type 
of charge order at ${\bf q}_{1/4}$ (${\bf q}_{\rm 1/2-I}$) 
around $\widetilde T=$0.04 (0.045) for 
$V_{\rm cNNN}/V_{\rm abNN}=0.58$ (0.62). 
Temperature dependence of $P$ 
at several values of $V_{\rm cNNN}/V_{\rm abNN}$ 
is presented in Fig.~\ref{fig:polarization}.
At $V_{\rm cNNN}/V_{\rm abNN}=0.6$, 
$P$ remains down to the low temperature limit.
Apart from $V_{\rm cNNN}/V_{\rm abNN}=0.6$, $P$ starts to decrease 
at the temperature where $N({\bf q}_{1/2\textrm{-II}})$ and $N({\bf q}_{1/4})$ 
grow up, and disappears at the lowest temperature.
These results obtained by the MUMC method 
are qualitatively consistent with the ones in the mean-field calculation.

The polar charge structure characterized 
by ${\bf q}_{1/3}$ and 
the transition to the another structure characterized by ${\bf q}_{1/4}$ 
at $V_{\rm cNNN}/V_{\rm abNN} < 0.6$ are consistent with the experimental results. 
In LuFe$_{2}$O$_{4}$, 
charge order indexed as $(1/3\ 1/3\ 3m+1/2)$ appears around 350K 
and remains, at least, down to around 20K.~\cite{yamada00} 
On the other hand, 
in YFe$_{2}$O$_{4}$,  
charge order indexed as $(1/3\ 1/3\ 3m+1/2)$ observed at room temperature 
is changed into the one as 
$(1/4\ 1/4\ 3/4)$ around 250K.~\cite{ikeda02,ikeda03}
We suppose that different rare-earth metal ions 
slightly change ratio of the Coulomb potentials,  
and LuFe$_{2}$O$_{4}$ (YFe$_{2}$O$_{4}$) corresponds to 
the parameter region of $V_{\rm cNNN}/V_{\rm abNN} \thickapprox 0.6$ 
($V_{\rm cNNN}/V_{\rm abNN} < 0.6$) in the present calculation. 

Stability of the CO$_{1/3}$ phase is attributed to 
large thermal fluctuation in the geometrically frustrated lattice. 
A key issue is the two-sublattice structure in this charge ordered phase  
(see Fig.~\ref{fig:MF1/3}):~\cite{two_sub} 
Fe$^{2+}$ ions in the 2Fe$^{2+}$-Fe$^{3+}$ (lower) plane  
and Fe$^{3+}$ in the Fe$^{2+}$-2Fe$^{3+}$ (upper) one belong to 
a sublattice termed A. 
Other Fe ions belong to another sublattice termed B. 
All in-plane NN ions of a site in the sublattice B 
have an opposite valence. 
On the other hand, a site on the sublattice A is surrounded by three NN Fe$^{2+}$ 
and three NN Fe$^{3+}$ in the plane. 
Therefore, the Coulomb potentials at these sites from the in-plane NN ions are canceled out, 
and charge fluctuation is able to occur easily without loss of $V_{\rm abNN}$. 
It is obtained in the numerical calculation that 
an amplitude of the mean-field on the sublattice A is 0.9$V_{\rm abNN}$ at low temperature 
which is much less than that on the sublattice B, $2.1V_{\rm abNN}$. 
Large charge fluctuation at the sites grows up with increasing temperature, 
and contributes to the entropy gain at finite temperature. 
On the contrary, in the CO$_{1/2}\textrm{-II}$ and CO$_{1/4}$ structures, 
all Fe$^{2+}$ (Fe$^{3+}$) are equivalent
and charge fluctuation is weaker than that in the sublattice A of CO$_{1/3}$.
This is the reason why the polar charge order characterized by  
$(1/3, \ 1/3, \ 0)$ is more stable than other charge structures in finite temperatures. 

Let us focus on the charge structure in low temperatures in more detail. 
As shown in Figs.~\ref{fig:cchv} (b) and \ref{fig:polarization}, 
saturated values of $N({\bf q}_{1/3})$ and $P$ in $V_{\rm cNNN}/V_{\rm abNN}$=0.6
at the low temperature limit are 0.032 and 0.094, respectively, 
which are smaller than the values  
expected from the ideal CO$_{1/3}$ phase, 0.056 and 0.33 respectively.
This implies that the charge configuration at low temperature in  
$V_{\rm cNNN}/V_{\rm abNN}=0.6$ is not the ideal CO$_{1/3}$ state.
We analyze the probability histogram in the MUMC simulation, 
and examine the charge configurations realized in lowest temperatures.
These are classified into the following three configurations: 
the polar CO$_{1/3}$ structure shown in Fig.~\ref{fig:COs}~(a), 
partially polarized charge structures characterized by 
the momentum ${\bf q}_{1/3}$, termed CO$_{\rm A}$, 
and non-polar structures termed CO$_{\rm B}$.
Detailed structures of CO$_{\rm A}$ 
and CO$_{\rm B}$ are shown in Appendix \ref{sect:ap2}.
In CO$_{\rm A}$, the polarization is $P=N/3-n \sqrt{N}$ 
with an integer number $n$ satisfying a relation of $0 \leq n \leq 2\sqrt{N}/3$. 
Degeneracy of a sum of these configurations is 
of the order of 
$\sum_n \ _{2\sqrt{N}/3} {\rm C}_n \sim 2^{\sqrt{N}}$. 
As for the CO$_{\rm B}$ state, 
degeneracy of the configuration is also of the order of $2^{\sqrt{N}}$.
Because of the coexistence of these charge structures, 
the saturated values of $P$ and $N({\bf q}_{1/3})$ are smaller 
than the expected values from the ideal CO$_{1/3}$ state.
This tendency is remarkable in the large system size, as shown in Fig.~\ref{fig:polarization}. 

This coexistence of the polar and non-polar CO states 
implies that the full polarization expected from the ideal CO$_{1/3}$ state 
is realized by additional weak interactions. 
The long-range Coulomb interactions 
between the NN W-layers is one of the candidates. 
This scenario is plausible, since, 
in LuFe$_{2}$O$_{4}$, 
the electric polarization appears 
around the three-dimensional charge-ordering temperature.~\cite{ikeda05,ikeda00}
We examine effects of the inter W-layer Coulomb interaction 
based on a model where two W-layers stacked along the $c$ axis 
are coupled by the Coulomb interaction. 
Saturated values of $N({\bf q}_{1/3})$ and $P$ at low temperature 
are identical to the expected values from 
the ideal polar CO$_{1/3}$ state.
Roles of the exchange interaction as another candidate to lift the degeneracy  
are examined in the next section. 

\section{Spin Structure and Magneto-Electric effect}
\label{sect:spin}

In this section, we introduce spin degree of freedom and 
examine coupling between the electric polarization 
and the magnetic ordering.
The Hamiltonian ${\cal H}_{V}+{\cal H}_{J}$ is analyzed by utilizing the MUMC method 
in a 6$\times$6$\times$2-site cluster. 
The spin operators ${\bf I}_i$ and ${\bf J}_i$ in ${\cal H}_{J}$ are 
assumed to be Ising spins because of the strong magnetic anisotropy 
observed in $R$Fe$_{2}$O$_{4}$.~\cite{iida93}
The energy parameters in the Hamiltonian are chosen to be 
$U^{d}=7.8$, $W^{d}=6.2$, $I^{d}=0.8$, 
$U^{p}=4.1$, $W^{p}=2.9$, $I^{p}=0.6$, 
$t^{pd}=1.8$, $\Delta_{\rm CT}=3$ and $\widehat V=1$
in a unit of $V_{\rm abNN}$. 
These are determined from the experimental date in LaFeO$_3$.~\cite{mizokawa96,mizuno98} 
In this section, the orbital pseudo-spin operators in ${\cal H}_J$ 
are set to be zero, as explained in Sect.~\ref{sect:model}.
In particular, we focus on a parameter region around $V_{\rm cNNN}/V_{\rm abNN}=0.6$, 
where CO$_{1/3}$ is seen down to the lowest temperature in Fig.~\ref{fig:COphaseT}, 
and that around 0.58-0.59, where the transition from 
CO$_{1/3}$ to CO$_{\rm 1/4}$ is shown in Fig.~\ref{fig:COphaseT}. 

\begin{figure}
\centering
\includegraphics[width=0.8\columnwidth]{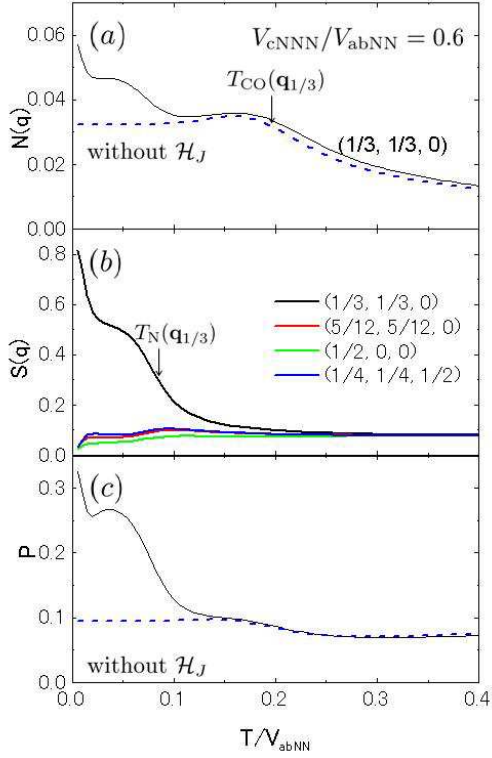}
\caption{
Charge correlation functions N({\bf q}) (a),  
spin correlation functions $S({\bf q})$ (b), 
and electric polarization $P$ (c) 
calculated in ${\cal H}_{V}+{\cal H}_{J}$.
Dashed lines in (a) and (c) are results obtained in ${\cal H}_{V}$.
Parameters are chosen to be 
$V_{\rm cNN}/V_{\rm abNN}=1.2$ and 
$V_{\rm cNNN}/V_{\rm abNN}=0.60$. 
}
\label{fig:hj60}
\end{figure}
%
\begin{figure}
\centering
\includegraphics[width=0.8\columnwidth]{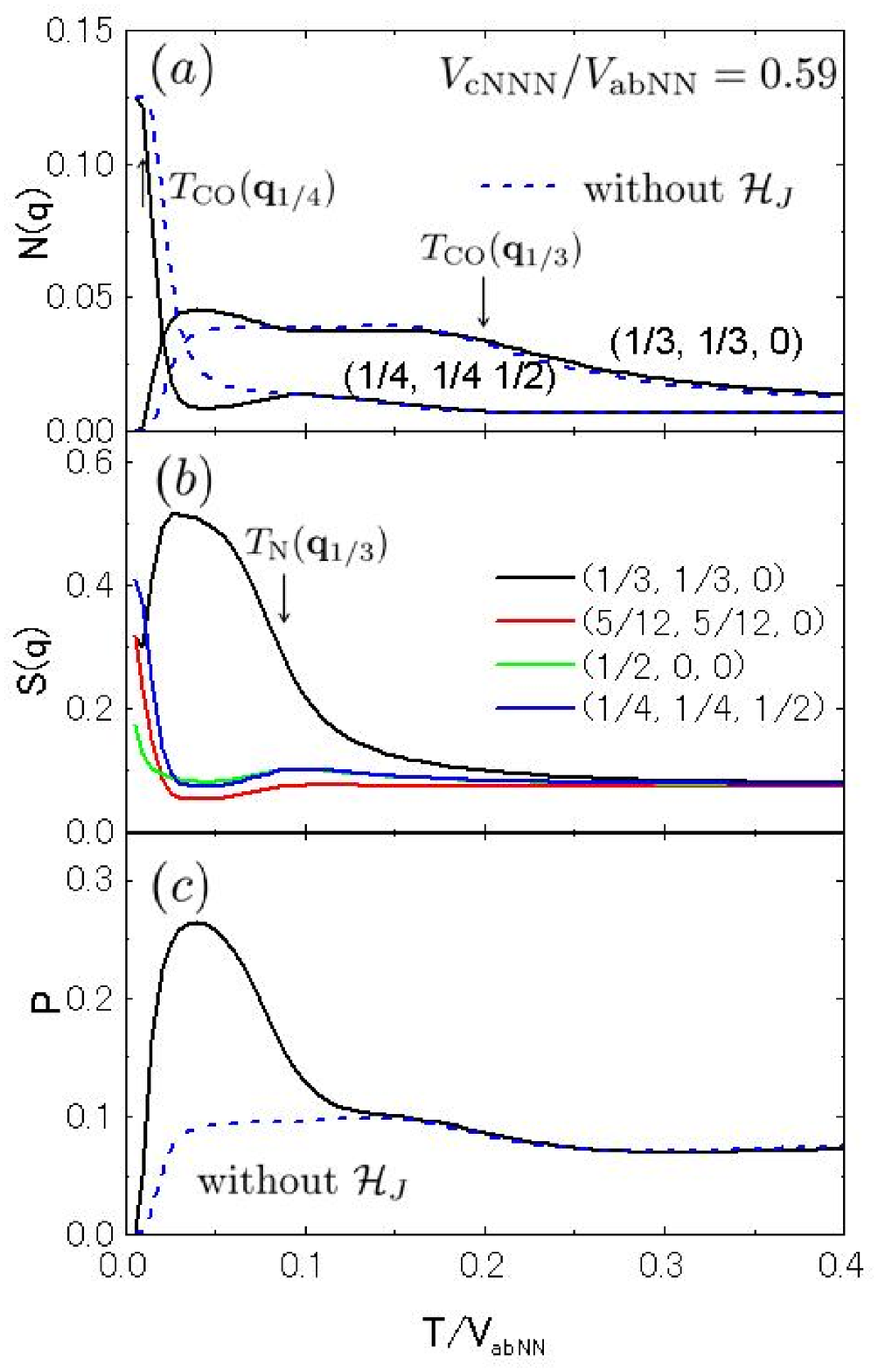}
\caption{
Charge correlation functions $N({\bf q})$ (a),  
spin correlation functions $S({\bf q})$ (b), 
and electric polarization $P$ (c) 
calculated in ${\cal H}_{V}+{\cal H}_{J}$.
Dashed lines in (a) and (c) are results obtained in ${\cal H}_{V}$.
Parameters are chosen to be 
$V_{\rm cNN}/V_{\rm abNN}=1.2$ and 
$V_{\rm cNNN}/V_{\rm abNN}=0.59$. 
}
\label{fig:hj59}
\end{figure}

Temperature dependences of 
the charge correlation function, 
the spin correlation function defined by
\begin{equation}
S({\bf q}) = \frac{1}{(2N)^2} \sum_{ij} 
\langle K_{i}^z K_{j}^z \rangle e^{-i {\bf q} \cdot ({\bf r}_{i}-{\bf r}_{j})} , 
\end{equation}
where $K_{i}^z=I_{i}^z$ $(J_{i}^z)$ for Fe$^{2+}$ (Fe$^{3+}$), 
and the electric polarization are calculated. 
Results at $V_{\rm cNNN}/V_{\rm abNN}=0.60$ and 0.59 
are shown in Figs.~\ref{fig:hj60} and \ref{fig:hj59}, respectively. 
For comparison, we also plot the data obtained in ${\cal H}_{V}$. 
At $V_{\rm cNNN}/V_{\rm abNN}=0.6$, 
three characteristic temperatures, $\widetilde T=0.2$, $0.085$ and $0.015$, are seen in $N({\bf q}_{1/3})$.  
The highest one, 0.2$ [\equiv T_{\rm CO}({\bf q}_{1/3})]$, corresponds to the charge 
ordering temperature for CO$_{1/3}$. 
Other two are the magnetic ordering ones  
at momentum ${\bf q}_{1/3}$. 
At $\widetilde T=0.085 [\equiv T_{\rm N}({\bf q}_{1/3}) ]$ and 0.015, 
spins in the Fe$^{2+}$-2Fe$^{3+}$ and 
2Fe$^{2+}$-Fe$^{3+}$ planes in CO$_{1/3}$ (see Fig.~\ref{fig:COs}) start to order, respectively. 
This double-magnetic transition may be an artifact 
in the present model where the inter-plane exchange interactions are neglected, 
and spins in the upper and lower planes are independent with each other. 
We expect that the inter-plane exchange interactions are much smaller than the in-plane ones.  
This is because, when electrons in the $d_{xy}$ and $d_{x^2-y^2}$ orbitals are concerned, 
there are no exchange paths between Fe ions in an inter-plane NN bond. 
When higher-order exchange processes and/or contributions from other $d$ orbitals are taken into account, 
weak inter-plane interactions may unify the double transition in the present calculation. 
As shown in Fig.~\ref{fig:hj60}, 
the charge correlation function at ${\bf q}_{1/3}$ and 
the polarization increase at $\widetilde T=0.085$ and $0.014$.
Results clearly show that magnetic ordering enhances 
stability of the polar CO$_{1/3}$ phase.  
In the low temperature limit, $N({\bf q}_{1/3})$ and $P$ take 
$0.056$ and $0.33$ respectively, which are the ideal values in CO$_{1/3}$.
At $V_{\rm cNNN}/V_{\rm abNN}=0.59$ (Fig.~\ref{fig:hj59}), 
a weak shoulder in $N({\bf q}_{1/3})$ around $\widetilde T=0.2$ 
corresponds to the charge ordering for CO$_{1/3}$. 
Sequential charge ordering transition occurs  
from CO$_{1/3}$ to CO$_{1/4}$ around $\widetilde T=0.015$  $[\equiv T_{\rm CO} ({\bf q}_{1/4})]$,  
which is lower a little than the result in ${\cal H}_V$. 
Magnetic order at ${\bf q}_{1/3}$ appears around  $\widetilde T=0.1 [\equiv T_N({\bf q}_{1/3})]$. 
Below $T_{\rm CO} ({\bf q}_{1/4})$,  
magnetic structure is also changed; the spin correlation functions at 
${\bf q}_{\rm 1/4}$ and $(5/12,\ 5/12,\ 0)$ 
become dominant. 
It is also shown, in this parameter, that 
the electric polarization is enhanced in the CO$_{1/3}$ phase. 
A similar temperature dependence is obtained in $V_{\rm abNN}/V_{\rm cNNN}=0.61$, 
where the CO$_{1/2}\textrm{-II}$ phase appears  
in low temperatures instead of CO$_{1/4}$.

\begin{figure}
\includegraphics[width=0.9\columnwidth]{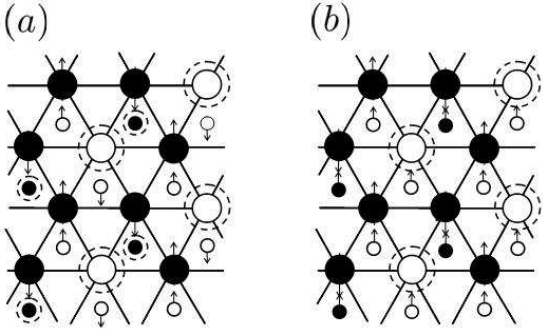}
\caption{
Charge and spin structures in the polar CO$_{1/3}$ phase at $V_{\rm cNNN}/V_{\rm abNN}=0.6$. 
Filled and open circles represent 
Fe$^{3+}$ and Fe$^{2+}$, respectively, 
and large and small circles are for Fe ions in the upper and lower planes, 
respectively.
Arrows represent spin directions.
At Fe sites surrounded by broken circles,  
spin directions are not determined uniquely.
Spin structures in (a) and (b) are almost degenerate.
}
\label{fig:spin3}
\end{figure}
\begin{figure}
\includegraphics[width=0.9\columnwidth]{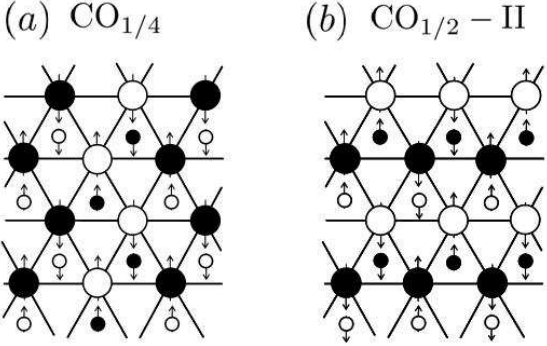}
\caption{
Charge and spin structure at $V_{\rm cNNN}/V_{\rm abNN}=0.59$ (a),  
and that at $0.61$ (b).
}
\label{fig:spin24}
\end{figure}

Low temperature charge and spin structures at $V_{\rm cNNN}/V_{\rm abNN}$=0.6  
are shown in Fig.~\ref{fig:spin3}(a). 
Charge structure is identified to be CO$_{1/3}$.
Spins at Fe$^{3+}$ in the Fe$^{2+}$-2Fe$^{3+}$ (upper) plane   
and those at Fe$^{2+}$ in the 2Fe$^{2+}$-Fe$^{3+}$ (lower) one 
are aligned antiferromagnetically.
On the other hand, 
spin directions at Fe$^{2+}$ in the Fe$^{2+}$-2Fe$^{3+}$ plane  
and at Fe$^{3+}$ in the 2Fe$^{2+}$-Fe$^{3+}$ one 
are not determined uniquely.
We note that the spin structure in the Fe$^{3+}$-2Fe$^{2+}$ (lower) plane  
is sensitive to the parameter values in ${\cal H}_{J}$. 
The structures shown in Figs.~\ref{fig:spin3}(a) and (b)  
are almost degenerate with each other. 
The numerical results presented in this paper are obtained in 
the parameter sets where the spin structure in Fig.~\ref{fig:spin3}(a) is obtained. 
However, qualitative difference for the results in the two parameter sets is not seen. 
It is also true that essence of the coupling between the spin ordering 
and the electric polarization shown in Fig.~\ref{fig:hj60} 
does not depend on the detailed parameter values.
Since the antiferromagnetic interaction between NN Fe$^{3+}$-Fe$^{3+}$ bonds 
in the 2Fe$^{3+}$-Fe$^{2+}$ (upper) plane is robust, 
Fe$^{2+}$ spins are surrounded by three up and three down spins 
in their NN Fe$^{3+}$ sites. 
Therefore, spin directions in Fe$^{2+}$ are not determined uniquely as explained above. 
Because the number of these sites is $N/3$, 
there is a macroscopic number of degenerate spin states 
of the order of $2^{N/3}$ 
which contributes to the entropy gain in finite temperatures. 
This is a kind of partially disordered phase, which has been examined in the antiferromagnetic 
Ising model on a triangular lattice.~\cite{metcalf74}
In the present case, 
spins in Fe$^{2+}$ and Fe$^{3+}$ are inequivalent, i.e. $S=2$ and $5/2$, 
and this partial disordered state becomes more stable in comparison with that in the conventional Ising model. 
Since this spin structure is realized in the CO$_{1/3}$ structure  
and the spin entropy is larger than the charge entropy 
in the non-polar and partially-polar charge ordered phases, i.e. CO$_{\rm A}$ and CO$_{\rm B}$, 
the polar CO$_{1/3}$ is reinforced through the spin-charge coupling 
in the exchange Hamiltonian. 
This is a kind of "order from fluctuation" mechanism, 
and, in the present spin-charge coupled system, 
a ferroelectric order is stabilized by spin fluctuation. 
This phenomenon is not expected in CO$_{\rm 1/2}$-I, CO$_{\rm 1/2}$-II and CO$_{1/4}$. 
Low temperature charge and spin structures in $V_{\rm cNNN}/V_{\rm abNN}=0.59$ and $0.61$ 
are shown in Fig.~\ref{fig:spin24}.
In both cases, 
all spins in NN Fe$^{2+}$-Fe$^{2+}$ and Fe$^{3+}$-Fe$^{3+}$ bonds 
are aligned antiferromagnetically. 
There are a number of degenerate spin states; for example, 
when all spins on a chain along [110] in CO$_{\rm 1/2}$-II 
are flipped, the exchange energy is not changed. 
However, this spin degeneracy is of the order of $2^{\sqrt{N}}$,  
which is smaller than $O(2^{N/3})$ in CO$_{1/3}$. 

\begin{figure}
\centering
\includegraphics[width=0.8\columnwidth]{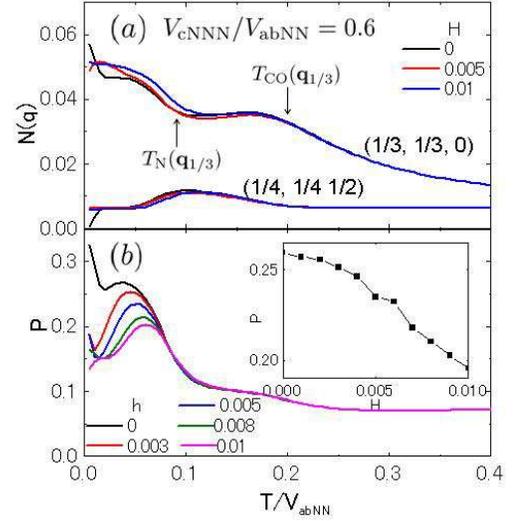}
\caption{
Magnetic-field effect in charge correlation function $N({\bf q})$ (a), 
and electric polarization $P$ (b).  
Parameters are chosen to be 
$V_{\rm cNN}/V_{\rm abNN}=1.2$ and $V_{\rm cNNN}/V_{\rm abNN}=0.6$.
Inset of (b) shows magnetic-field dependence of the electric polarization 
at $\widetilde T=$0.05. 
}
\label{fig:ME60}
\end{figure}
\begin{figure}
\centering
\includegraphics[width=0.8\columnwidth]{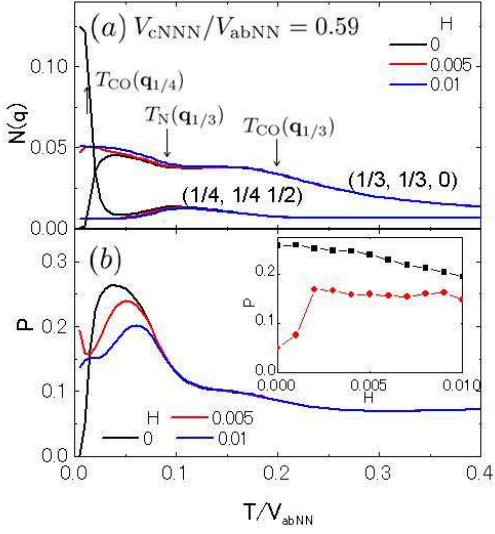}
\caption{
Magnetic-field effect in charge correlation function $N({\bf q})$ (a), 
and electric polarization $P$ (b).  
Parameters are chosen to be 
$V_{\rm cNN}/V_{\rm abNN}=1.2$ and $V_{\rm cNNN}/V_{\rm abNN}=0.59$.
Inset of (b) shows magnetic-field dependence of the electric polarization. 
Solid circles and squares are calculated at $\widetilde T=$0.01 and 0.05, respectively.  
}
\label{fig:ME59}
\end{figure}
In a remaining part of this section, 
we examine responses to the electric and magnetic fields  
in the present novel spin-charge coupled system.  
First we pay our attention to the magnetic-field effect  
by introducing the Zeeman term of the Hamiltonian  
\begin{equation}
{\cal H}_{H}=H \sum_{i} K^z_{i}, 
\end{equation}
where $K_{i}^z=I_{i}^z$ or $J_{i}^z$ 
for Fe$^{2+}$ or Fe$^{3+}$, respectively, and $H$ is the magnetic field. 
The Hamiltonian ${\cal H}_{\rm V}+{\cal H}_{J}+{\cal H}_{H}$
is analyzed by utilizing the MUMC method in a 6$\times$6$\times$2-site cluster.
Magnetic field dependence of 
the electric polarization 
and the charge correlation functions at 
$V_{\rm cNNN}/V_{\rm abNN}=0.6$ and 0.59 are presented 
in Figs.~\ref{fig:ME60} and \ref{fig:ME59}, respectively. 
Temperature in Fig.~\ref{fig:ME60} is chosen to be 
$\widetilde T=$0.05, which is below the N${\rm \acute e}$el temperature $T_{\rm N}({\bf q}_{1/3})$, 
and those in Fig.~\ref{fig:ME59} are
$\widetilde T=$0.05 and 0.01,
which are between $T_{\rm N}({\bf q}_{1/3})$ 
and the charge ordering temperature of CO$_{1/4}$ $[T_{\rm CO}({\bf q}_{1/4})]$, 
and below $T_{\rm CO}({\bf q}_{1/4})$, respectively. 
When $V_{\rm abNN}$ is taken to be 1eV, magnetic field 
$H/V_{\rm abNN}=0.01$ corresponds to about $100$Tesla.
In the magnetically ordered CO$_{1/3}$ phases 
at $V_{\rm cNNN}/V_{\rm abNN}$=0.6 and 0.59, 
applying the magnetic field reduces the electric polarization. 
On the other hand, 
in the antiferromagnetic CO$_{1/4}$ phase at $V_{\rm cNNN}/V_{\rm abNN}=0.59$ 
[see $T<T_{\rm CO}({\bf q}_{1/4})$ in Fig.~\ref{fig:ME59}(b)], 
the electric polarization is induced by applying the magnetic field. 
At the same time, 
the charge correlation function $N({\bf q}_{1/3})$ increases and 
$N({\bf q}_{\rm 1/4})$ decreases. 
Similar results are obtained at $V_{\rm cNNN}/V_{\rm abNN}=0.61$,  
where the CO$_{\rm 1/2-II}$ phase collapses and the electric polarization appears below 
$\widetilde T=0.015$ by applying the magnetic field.
Thus,  
opposite magnetic-field effects  
are obtained in the magnetically ordered CO$_{1/3}$ phase and the antiferromagnetic CO$_{1/4}$ 
and CO$_{\rm 1/2-II}$. 

We, first, pay our attention to the negative magnetic-field effect in 
the magnetically ordered CO$_{1/3}$ phase. 
As explained in Sect.~\ref{sect:charge}, 
the three charge structures, 
the polar CO$_{1/3}$, 
the partially polar CO$_{\rm A}$  
and the non-polar CO$_{\rm B}$, 
coexist at $H=0$. 
Among the three, the polar CO$_{1/3}$ is a dominant structure, 
because of the large spin entropy due to the spin degeneracy 
of the order of $2^{N/3}$.
By applying the magnetic field, 
these $N/3$ spins are aligned to be parallel to the magnetic field, 
and the macroscopic spin degeneracy is lifted.
On the other hand, 
in both the non-polar CO$_{\rm B}$ and the partially polar CO$_{\rm A}$, 
a macroscopic degeneracy in the charge configuration, which is 
of order of $2^{\sqrt{N}}$, 
survives under the magnetic field.
As the result, the charge entropy in CO$_{\rm A}$ and CO$_{\rm B}$  
overcomes the spin one in CO$_{1/3}$, 
and $P$ is reduced. 
In other words, 
under the magnetic field, 
the present spin-charge coupled system is mapped onto a spin-less model  
described by ${\cal H}_V$ where 
the charge entropy plays a dominant role. 
On the contrary, 
the positive magnetic-field effect 
in the antiferromagnetic CO$_{1/4}$ phase 
is explained from the Zeeman energy. 
Under the magnetic field, the ferrimagnetic structure in the polar CO$_{1/3}$ phase 
is more stable than the antiferromagnetic one in CO$_{1/4}$, 
and the polarization appears below $T_{\rm CO}({\bf q}_{1/4})$. 
However, under a high magnetic field larger than $H/V_{\rm abNN} \sim 0.01$, 
the polar CO$_{1/3}$ competes with CO$_{\rm A}$ and CO$_{\rm B}$, 
and the polarization decreases, as discussed above. 

\begin{figure}
\centering
\includegraphics[width=0.8\columnwidth]{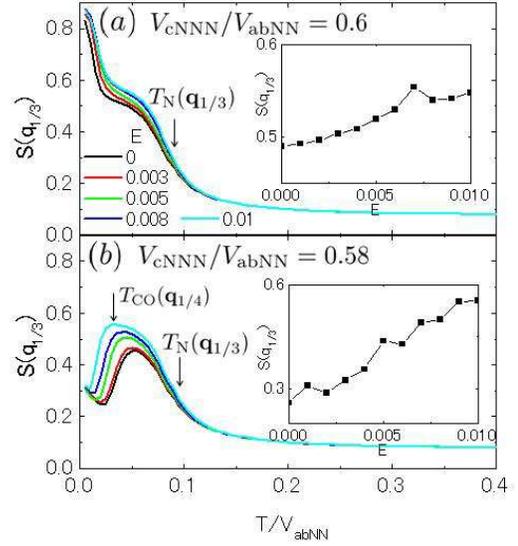}
\caption{
Electric-field effect of 
spin correlation function $S({\bf q})$ at ${\bf q}_{1/3}$.      
Parameters are chosen to be
$V_{\rm cNNN}/V_{\rm abNN}=0.6$ (a) and 0.58 (b) with $V_{\rm cNN}/V_{\rm abNN}=1.2$. 
Insets show electric-field dependence of $S({\bf q_{1/3}})$. 
Temperatures are $\widetilde T=$0.05 in (a), and 0.03 in (b). 
}
\label{fig:electric}
\end{figure}

We are also able to demonstrate the response to the electric field 
in the present spin-charge coupled system. 
The static electric field $E$ applied along the $c$ axis 
is formulated by the Hamiltonian 
\begin{eqnarray}
{\cal H}_E=-eEd \sum_i^{u} Q_i^z, 
\label{eq:electric}
\end{eqnarray}
where $d$ is a distance between the two triangular-lattice planes in a W-layer. 
Amplitude of the electric field 
$eEd/V_{\rm abNN}=0.01$ corresponds to about 50MV/m, 
when we take $V_{\rm abNN}=$1eV and $d=2.2 {\rm \AA}$. 
Electric-field dependence of the spin correlation functions 
at $V_{\rm cNNN}/V_{\rm abNN}$=0.6 and 0.58 are 
presented in Fig.~\ref{fig:electric}. 
By applying the electric field, 
the spin correlation at ${\bf q}_{1/3}$ is enhanced, 
in particular, below $T_{\rm CO}({\bf q}_{1/4})$ in $V_{\rm cNNN}/V_{\rm abNN}$=0.58. 
This is a consequence of the polar CO$_{1/3}$ phase stabilized 
by the electric field. 
The results would be use as a test for the present scenario. 

\section{Effect of Oxygen Deficiency}
\label{sect:def}

It is well known that 
several dielectric and magnetic properties in 
$R$Fe$_2$O$_4$, e.g. charge and spin ordering temperatures, 
are extremely sensitive to the oxygen stoicheometry, 
denoted by $\delta$ in the formula $R$Fe$_2$O$_{4-\delta}$.~\cite{funahashi84,ikeda03,horibe07} 
Effects of the oxygen deficiency in this system 
are recognized as the impurity effects 
in charge-spin coupled system with geometrical frustration. 
Here we examine roles of oxygen deficiency 
on the magneto-electric phenomena. 
We simulate following two aspects of the oxygen deficiencies; 
1) charge imbalance between Fe$^{2+}$ and Fe$^{3+}$, 
which is introduced by the modified charge conservation relation as $N^{-1}\sum_{i} Q^z_i =-2\delta$, 
and 2) random electro-static potential around defect sites.  
This is modeled by the Hamiltonian 
\begin{equation}
{\cal H}_{\rm R}=2 \sum_{i} \sum_{j}^{\prime} V_{\rm R}(|i-j|) Q^z_{j}, 
\end{equation}
where $\sum_{i}$ and $\sum_j^\prime$ represent summations of defect sites 
and that of the neighboring Fe sites, respectively. 
We assume that a defect site is in the FeO plane, and effective charge of a defect is 2+. 
Amplitudes of the electro-static potentials are estimated by  
the $1/r$-type potential as 
$V_{\rm R}=1.73V_{\rm abNN}$ and $1.60V_{\rm abNN}$ 
for the NN and NNN sites from a defect site, respectively. 
The model Hamiltonian 
${\cal H}_{V}+{\cal H}_{J}+{\cal H}_{\rm R}$ 
is analyzed with the relation  $N^{-1} \sum_{i}Q_i^z=-2\delta$ 
by the MUMC method.
One defect is introduced in a $6 \times 6 \times 2$ site cluster.
This concentration corresponds to $\delta=0.05$.
%
\begin{figure}[t]
\centering
\includegraphics[width=0.8\columnwidth]{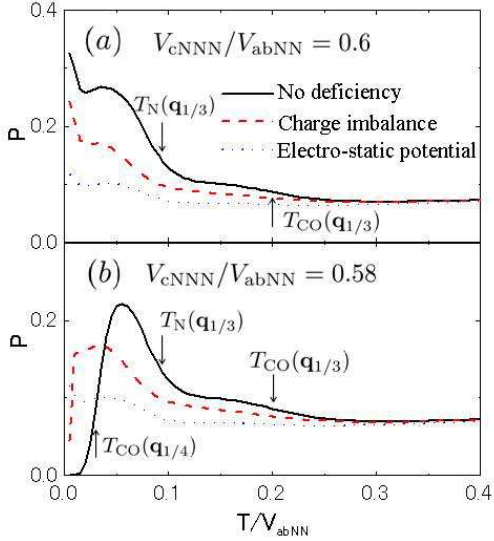}
\caption{
Oxygen-deficiency effects of 
electric polarization $P$  
at $V_{\rm cNNN}/V_{\rm abNN}=0.6$ (a), and 0.58 (b). 
Bold, broken and dotted lines are for result  
without deficiency, that with charge-imbalance effect, 
and that with electro-static potential, respectively. 
}
\label{fig:IMP60}
\end{figure}

In Fig.\ref{fig:IMP60}, 
oxygen-deficiency effect on  
the electric polarization 
is presented.
In CO$_{1/3}$ [see  
below $T_{\rm CO}({\rm q}_{1/3})$ in Fig.~\ref{fig:IMP60}(a) 
and between $T_{\rm CO}({\bf q}_{1/3})$ and $T_{\rm CO}({\bf q}_{1/4})$ in Fig.~\ref{fig:IMP60}(b)], 
both the charge-imbalance and electro-static potential effects 
suppress the electric polarization. 
On the contrary, 
in the antiferromagnetic CO$_{1/4}$ phase below $T_{\rm N}({\bf q}_{1/4})$ in Fig.~\ref{fig:IMP60}(b), 
the electric polarization is induced by both the two-types of deficiency effects.  
These results are consistent with the electron-diffraction experiments in YFe$_2$O$_{4-\delta}$;~\cite{horibe07} 
in samples with large $\delta$, 
the four fold-type charge order disappears, but  
the three fold-type indexed as $(1/3\ 1/3\ 3m+1/2)$ is robust. 

We turn to explain a mechanism of the charge-imbalance effect. 
Reduction of $P$ in the CO$_{1/3}$ phase shown in Fig.~\ref{fig:IMP60}(a) 
is a kind of an usual impurity effect which tends to break the long-range order. 
On the contrary, increase of $P$ in low temperatures shown in Fig.~\ref{fig:IMP60}(b) 
is related to the characteristic charge frustration in CO$_{1/3}$ as follows. 
The charge imbalance represented by a relation $N^{-1}\sum_{i} Q^z_i =-2\delta$ 
implies replacement of some Fe$^{3+}$ ions in a stoichiometric system by Fe$^{2+}$. 
This corresponds to flipping of pseudo-spins $Q_i^z$. 
It is rather trivial, in Fig.~\ref{fig:COs}, 
that this flipping of a $Q_i^z$ happens uniquely in the CO$_{1/2}$-I, CO$_{1/2}$-II and CO$_{1/4}$ structures. 
However, in CO$_{1/3}$, 
there are two ways to flip a $Q_i^z$ because of the two-sublattice 
structure mentioned in Sect.~\ref{sect:charge}: 
Fe$^{2+}$ sites surrounded by NN three Fe$^{2+}$ and three Fe$^{3+}$ in a plane 
(sublattice A), and those surrounded by six NN Fe$^{3+}$ in a plane (sublattice B). 
A pseudo-spin in the sublattice A is able to be flipped easily in energy.  
We numerically calculate energy costs due to a flipping in sublattice A 
is about 40$\%$ of that in  sublattice B, and is about 65$\%$ in CO$_{1/2}$-II and CO$_{1/4}$. 
Such low-energy charge excitations in CO$_{1/3}$ 
stabilize the charge structure under the charge imbalance. 

The electro-static potential effect is also understood from 
a viewpoint of a soft charge structure in the CO$_{1/3}$ phase. 
Since an effective charge of a defect is 2+, 
Fe$^{2+}$ ions, rather than Fe$^{3+}$, tend to assemble to screen this 
positive excess charge. 
However, due to the Coulombic interaction between Fe$^{2+}$ ions, 
a simple cluster consisting of Fe$^{2+}$ around a defect is not energetically favored. 
Exchange of Fe$^{2+}$ and Fe$^{3+}$ between the planes in a W-layer 
is able to reduce such Coulombic energy. 
Energy cost for this kind of exchange of Fe$^{2+}$ and Fe$^{3+}$ 
is much lower in CO$_{1/3}$ than that in other charge ordered structures. 
That is, 
the electro-static screening for excesses charge 
easily occurs in CO$_{1/3}$ because of the two-sublattice structure. 

\section{Discussion and Concluding Remark}
\label{sect:summary}

Here we have remarks on some issues which are not included explicitly in the present model and calculation. 
Effects of the electron transfer in 3$d$ orbitals 
are not taken into account in the Hamiltonian Eq.~(\ref{eq : effective Hamiltonian}). 
This may be reasonable for the first-step theoretical model in \rfeo. 
It is because, even above the three-dimensional charge-ordering temperature (250K) in YFe$_2$O$_4$, 
the electric resistivity $\rho$ shows an insulating behavior; 
$\rho$ increases with decreasing temperature.  
A magnitude of $\rho$ about 250K is of the order of 10$^2 \Omega$cm,~\cite{tanaka82} 
which is much larger that that above the Verwey transition in Fe$_3$O$_4$.~\cite{kakudate79} 
Therefore, we suppose that dominant electron motion is caused by thermal motion, rather than 
quantum electron transfer.  
This is supported by the experimental data in the dielectric constant; 
it shows strong dispersive feature  
well described by the Debye model based on the thermal fluctuation of dipole 
moments.~\cite{ikeda05,ikeda00,ikeda94b} 
Electron-transfer effects for the charge ordered phase in a triangular lattice have been 
investigated for some low-dimensional organic salts.~\cite{seo05,hotta06} 
In theoretical calculations based on the $V-t$ and extended Hubbard models at quarter filling, 
a metallic phase appears in a parameter region between 
two different charge orders, or 
it coexists with the three-fold type charge order.  
We suppose that small electron transfer in \rfeo \ 
stabilizes the CO$_{1/3}$ phase, 
although diffusive features in the dielectric function becomes more remarkable. 

Lattice degree of freedom and a coupling with electron are not 
included explicitly in the present calculation. 
In our knowledge, there are no detailed crystal structure data  
in spin-charge ordered phases. 
It is thought from the experimental analyses in YFe$_2$O$_4$ that 
the crystal symmetries in both the two- and three-dimensional 
charge-ordered phases indexed as $(1/3\ 1/3 \ 3m+1/2)$ are trigonal, 
but that in the four-fold type charge order is monoclinic.~\cite{horibe07} 
This result indicates that the lattice distortion in the 
three-fold type charge order is weaker than that in other phases.  
This is consistent with the present results for 
a soft charge-order character in CO$_{1/3}$; 
amplitude of the charge correlation function is smaller than that in other phases.   
A weak lattice distortion expected in the three-fold charge order 
is also related to the symmetry of the CO$_{1/3}$ structure,  
where the rhombohedral symmetry remains in a FeO planes, 
unlike other charge ordered phases. 

In Sect.~\ref{sect:spin}, 
We show that the electric polarization is reinforced by the ferrimagnetic ordering, 
and this originates from the spin entropy due to the frustrated geometry. 
The long-range exchange interactions and/or the magneto-striction effects,  
which are not included explicitly in the present model, 
may release the spin degeneracy.  
In these cases, we suppose that 
the spins, which are not fixed in the present model [see Fig.~\ref{fig:spin3}], 
are loosely bounded by such low-energy scale interactions. 
However, these still fluctuate in a temperature region which is higher than the energy scale of the interactions, 
and contribute to the entropy gain. 

In the present paper, we analyze an electronic model defined in a single W-layer 
which is recognized as a minimal and main stage in \rfeo.  
Obtained results provide a starting point 
to elucidate  a variety of magnetic and dielectric phenomena. 
We briefly discussed, in Sect.~\ref{sect:charge}, some roles of the inter W-layer Coulomb interaction.  
To clarify the three-dimensional charge and spin structures,~\cite{yamada00,funahashi84,christianson07,nagai07} 
and the magneto-dielectric response along the $c$ axis, 
a more realistic modeling for the inter W-layer interactions, in particular, the 
inter W-layer exchange interactions, and 
analyses of a three-dimensional model are necessary. 

In summary, electronic structure and magneto-dielectric phenomena in the 
rare-earth iron oxides with frustrated geometry are 
examined. 
The model Hamiltonian describing the electronic interactions between charge, spin and orbital 
degrees of freedom of Fe ions is derived. 
This model is analyzed by utilizing mainly the Monte-Carlo technique in a finite size cluster. 
The three fold-type charge order associated with electric polarization 
is stabilized in finite temperature in comparison with two and four fold-type charge structures. 
The two-sublattice structure in this polar charge order 
plays a crucial role. 
This phase is reinforced by the magnetic ordering, 
due to the spin frustration and the coupling between charge and spin in the exchange Hamiltonian. 
Novel magneto-electric responses to the external fields  
are available as a test of the present scenario. 
Effects of the oxygen deficiency  
are understood from the viewpoint of impurity effects in a frustrated spin-charge coupled system. 
Through the present study, we provide a unified scenario for a variety of 
magnetic and dielectric phenomena in \rfeo. 

\appendix
\section{Exchange Hamiltonian}
In this appendix, we show details of the superexchange processes 
and an explicit form of the Hamiltonian ${\cal H}_{J}$ 
introduced in Sect.~\ref{sect:model}.
There are two kinds of the superexchange processes 
termed the $dd$- and $dpd$-processes as introduced in Eqs.~(\ref{eq:dd}) and (\ref{eq:dpd}). 
The Hamiltonian ${\cal H}_{J}$ is classified by valences of Fe ions, 
Fe$^{m+}$-Fe$^{n+}$ in the initial 
and final states, and the electronic structure in the intermediate states $k$ [see Eq.~(\ref{eq:hj})]. 
In this appendix, nearest neighboring Fe sites concerning 
in the superexchange interactions are denoted as $i$ and $j$. 
Electron configuration in Fe and O ions are represented in a hole picture.

\subsection{Exchange Interactions in Fe$^{2+}$-Fe$^{2+}$}
%
\begin{figure}[t]
\includegraphics[width=0.9\columnwidth]{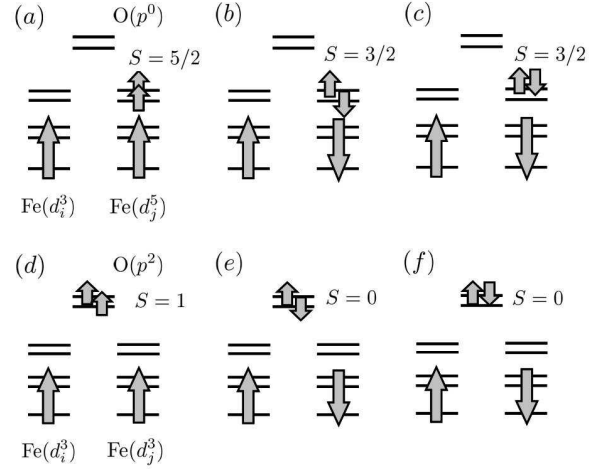}
\caption{
Intermediate states of the exchange processes in a Fe$^{2+}$-Fe$^{2+}$ bond 
represented by hole picture. 
(a), (b) and (c) are for the Hamiltonian ${\cal H}^{(22)-1}$, ${\cal H}^{(22)-2}$, and ${\cal H}^{(22)-3}$ 
in the $dd$-processes, respectively, 
and  (d), (e), and (f) are for ${\cal H}^{(22)-4}$, ${\cal H}^{(22)-5}$, and ${\cal H}^{(22)-6}$ 
in the $dpd$-ones, respectively. 
Long and short arrows represent spins 
with $S=3/2$ and $1/2$, respectively.
}
\label{fig:ex22}
\end{figure}
For the $dd$-processes, 
electron configurations in the intermediate exchange processes 
are denoted as $d^{3}p^{0}d^{5}$. 
Five holes are at a site $j$ 
and three holes at $i$ with $S=3/2$.
The intermediate states are classified by the spin and orbital states at site $j$
[see Fig.~\ref{fig:ex22}(a)-(c)]: 
(a) the total spin of Fe holes at site $j$, $S$, is equal to $5/2$ 
and both the two ${\rm E}^{\prime}$ orbitals are occupied, 
(b) $S=3/2$ and the two ${\rm E}^{\prime}$ are occupied, 
and (c) $S=3/2$ and one of the ${\rm E}^{\prime}$ is occupied.
The explicit forms of the exchange Hamiltonian are given by 
\begin{eqnarray}
{\cal H}^{(22)-1} &=& J^{(22)-1} 
\sum_{\langle ij \rangle} \left({\bf I}_{i} \cdot {\bf I}_{j} + 6 \right) 
\nonumber \\
&\times&
\left ( P^{\tau +}_i P_j^{\tau -} + P^{\tau -}_i P_j^{\tau +} \right ) 
P_i^{Q-} P_j^{Q-}, 
\end{eqnarray}
\begin{eqnarray}
{\cal H}^{(22)-2} &=& J^{(22)-2} 
\sum_{\langle ij \rangle} 
\left({\bf I}_{i} \cdot {\bf I}_{j} - 4 \right) 
\nonumber \\
&\times&
\left ( P^{\tau +}_i P_j^{\tau -} + P^{\tau -}_i P_j^{\tau +} \right ) 
P_i^{Q-} P_j^{Q-}, 
\end{eqnarray}
\begin{equation}
{\cal H}^{(22)-3} = J^{(22)-3} \sum_{\langle ij \rangle} 
\left({\bf I}_{i} \cdot {\bf I}_{j} - 4 \right) 
P_i^{\tau +} P_j^{\tau +} 
P_i^{Q-} P_j^{Q-}, 
\end{equation}
Here we define the projection operators for charge 
\begin{equation}
P_i^{Q \pm}= \frac{1}{2} \pm Q_i^z ,  
\end{equation}
and those for orbital 
\begin{equation}
P_i^{\tau \pm}= \frac{1}{2} \pm \tau_{i \eta_i} . 
\end{equation}
The exchange parameters are given as 
$J^{(22)-1}= -t_{dd{\rm c}}^{2}/[10 \Delta_{(22)-1}]$, 
$J^{(22)-2}=  t_{dd{\rm c}}^{2}/[10 \Delta_{(22)-2}]$, and 
$J^{(22)-3}=  t_{dd{\rm c}}^{2}/[4 \Delta_{(22)-3}]$,
and $\Delta_{(mn)-k}$ is the energy of the second order intermediate states 
given by 
$\Delta_{(22)-1} = W^d - I^{d}+ {\widehat V}$, 
$\Delta_{(22)-2} = W^d + 4I^{d}+ {\widehat V}$, 
and $\Delta_{(22)-3} = U^{d} + 4I^{d}+ {\widehat V}$.
We define $t_{dd{\rm c}}=t_{dd}^2 \cos \theta /\Delta_{\rm CT}$ 
and  $t_{dd{\rm s}}=t_{dd}^2 \sin \theta /\Delta_{\rm CT}$. 

In the intermediate states of the $dpd$-process, 
two holes occupy the O ion.
These states are classified 
by the spin and orbital states at the O site, 
[see Fig.~\ref{fig:ex22}(d)-(f)]: 
(d) the total spin of the O holes, $S$, is equal to $1$ 
and both the $p_x$ and $p_y$ orbitals are occupied by holes, 
(e) $S=0$ and two $p$ orbitals are occupied, 
and (f) $S=0$ and one of the $p$ orbitals occupied by holes.
The exchange Hamiltonians are given by 
\begin{equation}
{\cal H}^{(22)-4} = J^{(22)-4} 
\sum_{\langle ij \rangle} 
\left({\bf I}_{i} \cdot {\bf I}_{j} + 12 \right) 
P_{i}^{ \tau +} P_{j}^{ \tau +} 
P_{i}^{ Q-} P_{j}^{ Q-}, 
\end{equation}
\begin{equation}
{\cal H}^{(22)-5} = J^{(22)-5} \sum_{\langle ij \rangle} 
\left({\bf I}_{i} \cdot {\bf I}_{j} - 4 \right) 
P_{i}^{ \tau +} P_{j}^{ \tau +} 
P_{i}^{ Q-} P_{j}^{ Q-}, 
\end{equation}
\begin{equation}
{\cal H}^{(22)-6} = J^{(22)-6} \sum_{\langle ij \rangle} 
\left({\bf I}_{i} \cdot {\bf I}_{j} - 4 \right) 
P_{i}^{ \tau +} P_{j}^{ \tau +} 
P_{i}^{ Q-} P_{j}^{ Q-}. 
\end{equation}
The exchange parameters are given as 
$J^{(22)-4}=-t_{dd{\rm s}}^{2}/[4 \Delta_{(22)-4}] $, 
$J^{(22)-5}= t_{dd{\rm s}}^{2}/[4 \Delta_{(22)-5}] $, and 
$J^{(22)-6}= t_{dd{\rm c}}^{2}/[2 \Delta_{(22)-6}] $ 
with 
$\Delta_{(22)-4} = 2\Delta_{\rm CT} + W^p - I^{p}$, 
$\Delta_{(22)-5} = 2\Delta_{\rm CT} + W^p + I^{p}$, and 
$\Delta_{(22)-6} = 2\Delta_{\rm CT} + U^{p}$.

\subsection{Exchange Interactions in Fe$^{2+}$-Fe$^{3+}$}

\begin{figure}[t]
\includegraphics[width=0.9\columnwidth]{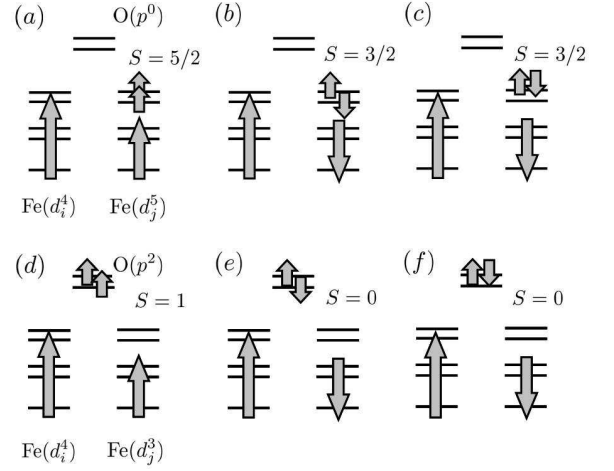}
\caption{
Intermediate states of the exchange processes in a Fe$^{2+}$-Fe$^{3+}$ bond 
represented by hole picture. 
(a), (b) and (c) are for the Hamiltonian ${\cal H}^{(23)-1}$, ${\cal H}^{(23)-2}$, and ${\cal H}^{(23)-3}$ 
in the $dd$-processes, respectively, 
and  (d), (e), and (f) are for ${\cal H}^{(23)-4}$, ${\cal H}^{(23)-5}$, and ${\cal H}^{(23)-6}$ 
in the $dpd$-ones, respectively. 
Long, medium and short arrows represent spins 
with $S=2$, $3/2$ and $1/2$, respectively.
}
\label{fig:ex23}
\end{figure}

Electron configurations in the intermediate states 
are $d^{4}p^{0}d^{5}$ and $d^{4}p^{2}d^{3}$ for the $dd$- and $dpd$-processes, 
respectively.
As well as the exchange interaction in the Fe$^{2+}$-Fe$^{2+}$ bond, 
these are classified by the spin and orbital structures 
in the $d^{5}$ and $p^{2}$ sites for the $dd$- and $dpd$-processes, 
respectively (see Fig.~\ref{fig:ex23}).
The explicit forms of the exchange Hamiltonians are  
\begin{equation}
{\cal H}^{(32)-1} = J^{(32)-1} 
\sum_{\langle ij \rangle} 
\left({\bf J}_{i} \cdot {\bf I}_{j} + \frac{15}{2} \right) 
P_j^{\tau -}
P_i^{Q +}P_j^{Q -}, 
\end{equation}
\begin{equation}
{\cal H}^{(32)-2} = J^{(32)-2} \sum_{\langle ij \rangle} \left({\bf J}_{i} \cdot {\bf I}_{j} - 5 \right) 
P_j^{\tau -} P_i^{Q +}P_j^{Q -}, 
\end{equation}
\begin{equation}
{\cal H}^{(32)-3} = J^{(32)-3} \sum_{\langle ij \rangle} \left({\bf J}_{i} \cdot {\bf I}_{j} - 5 \right) 
P_j^{\tau +} P_i^{Q +}P_j^{Q -}, 
\end{equation}
for the $dd$-processes, and 
\begin{equation}
{\cal H}^{(32)-4} = J^{(32)-4} \sum_{\langle ij \rangle} \left({\bf J}_{i} \cdot {\bf I}_{j} + 15 \right) 
P_j^{\tau +}
P_i^{Q +}P_j^{Q -}, 
\end{equation}
\begin{equation}
{\cal H}^{(32)-5} = J^{(32)-5} \sum_{\langle ij \rangle} \left({\bf J}_{i} \cdot {\bf I}_{j} - 5 \right)  
P_j^{\tau +}
P_i^{Q +}P_j^{Q -}, 
\end{equation}
\begin{equation}
{\cal H}^{(32)-6} = J^{(32)-6} \sum_{\langle ij \rangle} \left({\bf J}_{i} \cdot {\bf I}_{j} - 5 \right)  
P_j^{\tau +}P_i^{Q +}P_j^{Q -}, 
\end{equation}
for the $dpd$-ones. 
The exchange parameters are given as 
$J^{(32)-1}=-2 t_{dd{\rm c}}^{2}/[25\Delta_{(32)-1}]$, 
$J^{(32)-2}= 2 t_{dd{\rm c}}^{2}/[25\Delta_{(32)-2}]$,
$J^{(32)-3}=   t_{dd{\rm c}}^{2}/[10\Delta_{(32)-3}]$,
$J^{(32)-4}=-  t_{dd{\rm s}}^{2}/[5 \Delta_{(32)-4}]$, 
$J^{(32)-5}=   t_{dd{\rm s}}^{2}/[5 \Delta_{(32)-5}]$,  
and 
$J^{(32)-6}=2  t_{dd{\rm c}} ^{2}/[5\Delta_{(32)-6}]$ 
with 
$\Delta_{(32)-1} = {\widehat V}$, 
$\Delta_{(32)-2} = 5I^{d}+ {\widehat V}$, 
$\Delta_{(32)-3} = U^{d} - W^d + 4I^{d}+ {\widehat V}$, 
$\Delta_{(32)-4} = 2\Delta_{\rm CT} + W^p - I^{p}$, 
$\Delta_{(32)-5} = 2\Delta_{\rm CT} + W^p + I^{p}$, 
and $\Delta_{(32)-6} = 2\Delta_{\rm CT} + U^{p}$.

\subsection{Exchange Interactions in Fe$^{3+}$-Fe$^{3+}$}

\begin{figure}[t]
\includegraphics[width=0.9\columnwidth]{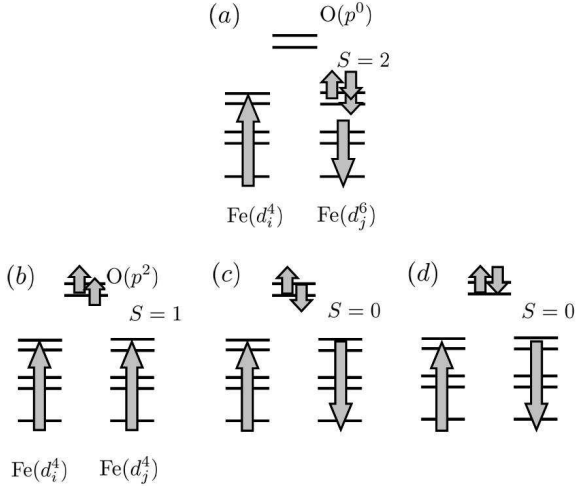}
\caption{
Intermediate states of the exchange processes in a Fe$^{3+}$-Fe$^{3+}$ bond 
represented by hole picture. 
(a) is for the Hamiltonian ${\cal H}^{(33)-1}$ in the $dd$-processes, respectively, 
and  (b), (c), and (d) are for ${\cal H}^{(33)-2}$, ${\cal H}^{(33)-3}$, and ${\cal H}^{(33)-4}$ 
in the $dpd$-ones, respectively. 
Long, medium and short arrows represent spins 
with $S=2$, 3/2 and $1/2$, respectively.
}
\label{fig:ex33}
\end{figure}
Electron configurations in the intermediate states 
are $d^{4}p^{0}d^{6}$ and $d^{4}p^{2}d^{4}$ for the $dd$- and $dpd$-processes, 
respectively.
In the $d^6$ configuration for the $dd$-process, 
total spin is 2 
[see Fig.~\ref{fig:ex33}(a)], 
and the explicit form is given by 
\begin{equation}
{\cal H}^{(33)-1} = J^{(33)-1} \sum_{\langle ij \rangle} 
\left({\bf J}_{i} \cdot {\bf J}_{j} - \frac{25}{4} \right)
P_i^{Q +}P_j^{Q +}. 
\end{equation}
The exchange parameter is 
$J^{(33)-1}=4 t_{dd{\rm c}}^{2}/[25\Delta_{(33)-1}]$
with $\Delta_{(33)-1} = U^{d} + 4I^{d}+{\widehat V}$.
For the $dpd$-processes, 
the intermediate states are classified by the spin and orbital structures 
in the O site [see Fig.~\ref{fig:ex33}(b)-(d)].
The Hamiltonians are given by 
\begin{equation}
{\cal H}^{(33)-2} = J^{(33)-2} 
\sum_{\langle ij \rangle} 
\left({\bf J}_{i} \cdot {\bf J}_{j} + \frac{75}{4} \right)
P_i^{Q +}P_j^{Q +}, 
\end{equation}
\begin{equation}
{\cal H}^{(33)-3} = J^{(33)-3} \sum_{\langle ij \rangle} \left({\bf J}_{i} \cdot {\bf J}_{j} - \frac{25}{4} \right)
P_i^{Q +}P_j^{Q +}, 
\end{equation}
\begin{equation}
{\cal H}^{(33)-4} = J^{(33)-4} \sum_{\langle ij \rangle} \left({\bf J}_{i} \cdot {\bf J}_{j} - \frac{25}{4} \right)
P_i^{Q +}P_j^{Q +}. 
\end{equation}
The exchange parameters are 
$J^{(33)-2}=-4 t_{dd{\rm s}}^{2} /[25 \Delta_{(33)-2}] $, 
$J^{(33)-3}= 4 t_{dd{\rm s}}^{2} /[25 \Delta_{(33)-3}] $ 
and 
$J^{(33)-4}= 8 t_{dd{\rm c}}^{2} /[25 \Delta_{(33)-4}] $ 
with 
$\Delta_{(33)-2} = 2\Delta_{\rm CT} + W^p - I^{p}$, 
$\Delta_{(33)-3} = 2\Delta_{\rm CT} + W^p + I^{p}$, 
and 
$\Delta_{(33)-4} = 2\Delta_{\rm CT} + U^{p}$.

\section{charge structures of CO$_{\rm A}$ and CO$_{\rm B}$}
\label{sect:ap2}

\begin{figure}[t]
\centering
\includegraphics[width=0.9\columnwidth]{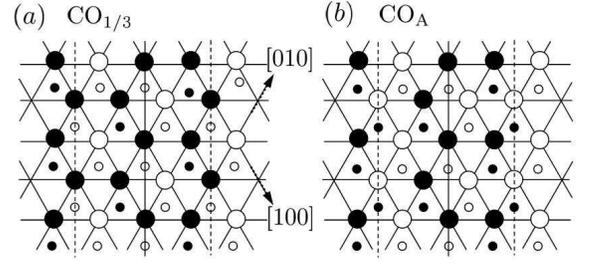}
\caption{
Schematic pictures of the CO$_{1/3}$ structure (a) and 
one of CO$_{\rm A}$(b). 
When, in CO$_{1/3}$, Fe$^{3+}$ in the upper plane and Fe$^{2+}$ in lower one 
on chains indicated by broken lines are exchanged, 
the CO$_{\rm A}$ structure in (b) is obtained. 
}
\label{fig:CO1/3-I}
\end{figure}
\begin{figure}[t]
\centering
\includegraphics[width=0.65\columnwidth]{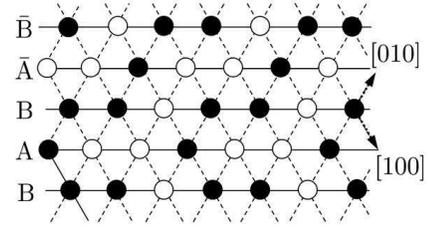}
\caption{
One of the CO$_{\rm B}$ structures. 
}
\label{fig:CO1/3-II}
\end{figure}

In this appendix, detailed charge structures in 
the CO$_{\rm A}$ and CO$_{\rm B}$ phases, 
introduced in Sect.~\ref{sect:charge}, are presented. 
Charge configurations of CO$_{\rm A}$ are constructed from CO$_{1/3}$.  
Start from the CO$_{1/3}$ structure shown in Fig.~\ref{fig:CO1/3-I}(a), 
and focus on chains, e.g. along $[ \bar{1}10 ]$,  
where different valences of Fe ions occupy the upper and lower planes. 
Let exchange all Fe$^{2+}$ and Fe$^{3+}$ in any of these chains with each other. 
One of the obtained configurations, termed CO$_{\rm A}$, is shown in Fig.~\ref{fig:CO1/3-I}(b). 
These structures of CO$_{\rm A}$ are energetically degenerate with CO$_{1/3}$ 
in the Hamiltonian ${\cal H}_V+{\cal H}_J$. 
The Coulomb interaction between the 2nd NN sites in the plane may lift the degeneracy. 
When the number of the chains, where Fe$^{2+}$ and Fe$^{3+}$ ions are exchanged, is 
$n$ $(0 \leq n < 2\sqrt{N}/3)$, 
the electric polarization is $P=N/3-n\sqrt{N}$. 
The degeneracy of a sum of these states is of the order of 
$\sum_n {}_{2 \sqrt{N}/3} {\rm C}_n \sim 2^{\sqrt{N}}$. 
Such exchange of Fe ions are also allowed on chains along $[ 120 ]$ and 
$[ 210 ]$ directions. 

In another degenerated structure, CO$_{\rm B}$, 
the configuration in one side of the W-layer 
is constructed by stacking two kinds of chains alternately. 
These chains are schematically given 
as $ \cdots \circ \circ \bullet \circ \circ \bullet \cdots$ 
(termed chain ${\cal A}$) 
and $\cdots \bullet \bullet \circ \bullet \bullet \circ \cdots$ 
(chain ${\cal B}$) 
along the $[ 110 ]$ direction 
where $\bullet$ and $\circ$ represent Fe$^{3+}$ and Fe$^{2+}$, respectively.  
As shown in Fig.~\ref{fig:CO1/3-II}, without energy loss of $V_{\rm abNN}$, 
there are two ways to stack a chain ${\cal A}$ on a chain ${\cal B}$, and vice versa. 
These are denoted as A and $\bar {\rm A}$, and B and $\bar {\rm B}$ in Fig.~\ref{fig:CO1/3-II}. 
Therefore, these configurations are degenerated of the order of $2^{\sqrt{N}}$.
Charge configuration on another side of the W-layer is uniquely determined 
to gain the inter-plane Coulomb interactions $V_{\rm cNN}$ and $V_{\rm cNNN}$. 
Obtained charge structures are degenerate with the CO$_{1/3}$ structure. 

\begin{acknowledgments}
The authors would like to thank 
N.~Ikeda, S.~Mori, T. Arima, Y. Horibe, J. Akimitsu, 
K.~Kakurai, N.~A.~Spaldin, M. Matsumoto, H. Matsueda and J. Nasu for their valuable discussions. 
This work was supported by JSPS KAKENHI (16104005), and 
 TOKUTEI gHigh Field Spin Science in 100T'' (18044001), 
gNovel States of Matter Induced by Frustration'' (19052001), and 
gInvention of Anomalous Quantum Materials''(19014003) from MEXT, 
NAREGI, and CREST. 
\end{acknowledgments}

$^{\ast}$ Present address: Japan Medical Materials Co., Osaka, 532-0003 Japan. 

\begin{thebibliography}{99}
\bibitem{two_sub}
When the Fe$^{2+}$ and Fe$^{3+}$ ions are distinguished, this structure has the four sublattices. 
However, we use, for simplicity, the two-sublattice structure.
\end{thebibliography}
\end{document}